\documentclass{IEEEtran}
\usepackage{cite}
\usepackage{xcolor}

\usepackage[final]{changes}
\makeatletter
\AddToHook{cmd/added/before}{\def\Changes@AuthorColor{magenta}}
\makeatother
\newcommand{\stkout}[1]{\ifmmode\text{\sout{\ensuremath{#1}}}\else\sout{#1}\fi}
\setdeletedmarkup{\stkout{#1}}
\usepackage{amsmath,amssymb,amsfonts}
\usepackage{bm}  
\usepackage{tikz}  
\usetikzlibrary{arrows.meta,backgrounds,decorations.pathreplacing,positioning,patterns,shapes.geometric, arrows,quotes,math}
\usepackage{subcaption}
\usepackage{caption}
\usepackage{algorithmic}
\usepackage{graphicx}
\usepackage{textcomp}
\usepackage{multirow}
\usepackage{booktabs}
\usepackage[normalem]{ulem}
 \def\BibTeX{{\rm B\kern-.05em{\sc i\kern-.025em b}\kern-.08em
    T\kern-.1667em\lower.7ex\hbox{E}\kern-.125emX}}
\begin{document}
\title{Topology Optimization of \\ Decoupling  Feeding Networks for Antenna Arrays}
\author{Pan Lu, Eddie Wadbro, Jonas Starck, Martin Berggren, and Emadeldeen Hassan
\thanks{This work was supported by the Swedish strategic research program eSSENCE and the Swedish Research Council grant 2018-03546. }
\thanks{P.\,Lu and M.\,Berggren are with the Department of Computing Science, Umeå University, 901\,87 Umeå, Sweden (e-mail: pan.lu@umu.se; martin.berggren@cs.umu.se). }
\thanks{E.\,Wadbro is with the Department of Mathematics and Computer Science, Karlstad University, 651\,88 Karlstad, Sweden, and the Department of Computing Science, Umeå University, 901\,87 Umeå, Sweden (e-mail: eddie.wadbro@kau.se).}
\thanks{J. Starck is with Proant AB, 906\,20 Umeå, Sweden, which was previously part of Abracon Corporation and is now a subsidiary of Mobile Mark Inc.(e-mail: jstarck@mobilemark.com).}
\thanks{E.\,Hassan is with the Department of Applied Physics and Electronics, Umeå University,  901\,87 Umeå, Sweden (e-mail: emadeldeen.hassan@umu.se).}}

\maketitle

\begin{abstract}
Near-field and radiation coupling between nearby radiating elements is unavoidable, and it is considered a limiting factor for applications in wireless communications and active sensing.
This article proposes a density-based topology optimization approach to design decoupling networks for such systems. 
The decoupling network is designed by formulating an optimization problem that considers both energy transmission and reflection at the network ports.
We replace the radiating elements by their time-domain impulse response for efficient computations and to enable the solution of the design problem using gradient-based optimization methods.
We use the adjoint-field method to compute the gradients of the optimization objectives.
Additionally, nonlinear filters are applied during the optimization procedure to impose minimum-size control on the optimized designs.
We demonstrate the concept by designing the decoupling network for a two-element planar antenna array; the antenna is designed in a separate optimization problem.
The optimized decoupling networks provide a signal path that destructively interferes with the coupling between the radiating elements while preserving their individual matching to the feeding ports.
Compact decoupling networks capable of suppressing the mutual coupling by more than 10\,dB between two closely separated planar antennas operating around 2.45\,GHz are presented and validated experimentally.

\end{abstract}

\begin{IEEEkeywords}
Antenna system, decoupling network, finite difference time domain (FDTD), impulse response boundary condition, topology optimization.
\end{IEEEkeywords}

\section{Introduction}
\label{sec:introduction}
\IEEEPARstart{M}{icrowave} systems consisting of multiple antennas for multiple-input-multiple-output (MIMO) communication are widely employed in modern wireless devices such as routers and mobile phones~\cite{jha2018compact,sun2018compact}. 
One well-known challenge in designing these compact systems is the mutual coupling between antennas or nearby transmission lines due to near-field interaction, which can deteriorate the overall system performance, including the signal-to-noise ratio, impedance matching, radiation pattern, and the overall radiation efficiency~\cite{sun2020self,hei2021wideband,pan2021design,liu2022mutual,lai2021mutual}. 
Therefore, mutual coupling reduction is a key aspect in the design of microwave components to ensure robust and high-performing systems.

Various decoupling techniques have been introduced in recent years to reduce mutual coupling in antenna arrays, including the use of additional circuits with microstrip lines and metallic components to enhance antenna isolation~\cite{li2021decoupling,zhang2021simple,pei2021low,zhao2024three,xu2020decoupling}, defected ground structures (DGSs) utilizing slots on the ground plane of planar circuits or antennas~\cite{khandelwal2017defected,qian2021decoupling}, neutralization lines to create a secondary path for the wave propagation through a narrow metallic structure~\cite{wang2013neutralization,zhang2015line}, metasurfaces consisting of periodic sub-wavelength elements to control wave interactions~\cite{Metasurface,Metasurface1}, or alterations of the antenna geometry~\cite{Liu24Full}.
The design of decoupling structures is a complex and time-consuming task, particularly when aiming for appropriate tradeoffs between various design objectives, including impedance matching, isolation, and size constraints. 
Optimization algorithms offer a systematic way to manage the complexity and address multiple design objectives including structure tolerances~\cite{Koziel13Multi,Hassan19Compact,Slawomir22Tolerance}.

The use of optimization algorithms to design structures for mutual coupling reduction is gaining interest and has the potential to revolutionize traditional design methodologies. 
An even-odd-mode and genetic algorithm based decoupling method is proposed by Cheng et al.~\cite{cheng2023novel} for a two inverted triangular antenna array. 
Genetic algorithms are also used by Zhang et al.~\cite{li2021decoupling} to design the connectivity of small metallic stubs inside a grid of decoupling networks for dualband MIMO antennas, which uses large blocks of metallic stubs between the microstrip lines. 
Although metaheuristic methods such as genetic algorithms are widely employed in microwave design problems, their applicability becomes limited for large-scale problems. 
This limitation arises because these methods use population-based search strategies and typically require two to three orders of magnitude more objective function evaluations than the number of design variables to reach satisfactory solutions~\cite{Hassan14patch,ko2024}.
Therefore, their use is limited to problems with predefined structures or when prior knowledge of the decoupling components is known.

Over the last decades, algorithmic-based design optimization has emerged as a powerful tool for large-scale optimizations over thousands of design variables for the design of microwave devices. 
Topology optimization is one of the most popular approaches to obtain a customized design based on specified objective functions. 
This method has proven to be effective in a wide range of application areas, such as mechanics, acoustics, optics, and fluid dynamics~\cite{Mousavi24Topology,Gedeon2023,Hassan:22,sigmund2013topology,Gedeon25Time}. 
Topology optimization approaches have also been used in the field of electromagnetics for the design of antennas and microwave devices~\cite{liu2016mom,hassan2018topology,zhu2019design,emad2020waveguide,topothin,Tucek23Density}. 
Gradient-based topology optimization algorithms based on continuous variables ranging between 0 and 1, known as ``material distribution'' or ``density-based'' methods, are one of the most commonly used approaches, offering fast convergence and less computational cost compared to gradient-free or metaheuristic methods such as genetic algorithms~\cite{smith2019gathin}. 
However, intermediate densities or isolated pixels sometimes occur in the final design when using such methods and may result in ambiguities in the interpretations of the results or difficulties in manufacturing.
To tackle this issue, design constraints can implicitly be imposed during the optimization to control the minimum feature size.
Also, hybrid optimization methods combining density-based topology optimization and level-set methods are developed to obtain designs with smooth boundaries and controlled feature sizes~\cite{wang2017antenna,emad2020waveguide}.
For instance, a hybrid topology optimization method have been used to design isolation structures at 5.8\,GHz for two rectangular microstrip patch antennas~\cite{zhu2019design}, where full simulations of the antenna system are performed using the finite element method in frequency domain.

We here propose a density-based topology optimization approach to design decoupling networks. 
Unlike conventional approaches, our method does not rely on predefined decoupling components or analytical assumptions, making it a fully algorithm-driven approach. 
The design problem is formulated as an optimization task to minimize reflection from and coupling between the feeding ports while maximizing the energy delivered to the antenna. 
The antennas are represented by their time-domain impulse responses, which results in an efficient solution to the optimization problem and allows us to use gradient-based optimization.
While minimizing the coupling energy is solely related to the feeding port signals, maximizing the energy radiation into free space must be considered to avoid energy dissipation in the decoupling structure.  
We use the finite-difference time-domain (FDTD) method for the full-wave performance analysis, and the computation of the derivatives needed for the optimization algorithm is carried out using the adjoint-field method. 
As the proposed optimization method is based on the impulse response, it might be applied to other interference scenarios without detailed knowledge of the antenna structure or the surrounding environment. 
Furthermore, the decoupling structure is developed using large-scale gradient-based topology optimization without any prior assumptions about the decoupling components, leading to a systematic and algorithm-driven approach.
The decoupling networks are optimized using nonlinear filters to enforce a minimum feature size on the designs.
We present novel designs of decoupling structures capable of suppressing the mutual coupling between two closely spaced antennas by more than 10 dB with little impact on individual antenna matching and radiation efficiency. 

\section{Setup and problem statement}
\begin{figure}
\centering
\includegraphics[trim = 0mm 0mm 0mm 0mm, clip ,width=0.8\columnwidth,draft=false]{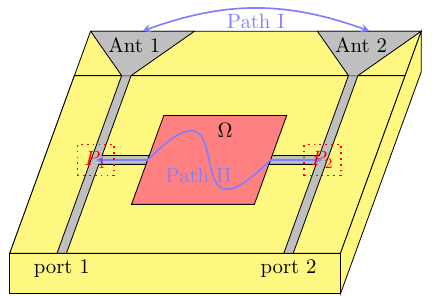}
\caption{Decoupling network design. The primary near-field coupling (Path\,I) between closely spaced antennas, Ant\,1 and Ant\,2, can be suppressed by optimizing a decoupling structure, in design domain $\Omega$, to create a secondary signal (Path\,II) that destructively interfere with the signal coupled through Path\,I.
}
\label{fig:prosetup}
\end{figure}

Fig.\,\ref{fig:prosetup} illustrates the conceptual setup of the design problem.
Two antennas are fed through two microstrip lines that are connected to port\,1 and port\,2, where incident/received signals are imposed/recorded. 
When the separation distance between the two antennas is small, which is typically required to achieve compact systems, a significant near-field mutual coupling can occur. 
To reduce this effect, a decoupling structure will be introduced within the design domain $\Omega$ located on top of a substrate and positioned between the two microstrip lines.  
The decoupling structure is connected to the microstrip lines through short microstrip lines possessing the same characteristic impedance as the main lines.
In the design domain $\Omega$, we aim to optimize the conductivity distribution $\sigma_\Omega$ of a good conductor (copper) to divert the original signal with an additional path to reduce the mutual coupling without significantly affecting the antenna performance. 
As illustrated in Fig.\,\ref{fig:prosetup}, there are two paths for the signal connecting port\,1 and port\,2, one traveling via path\,I in free space and the other through path\,II via the decoupling structure.  
With a properly designed domain $\Omega$, the signal traveling through path\,II could be processed to interfere destructively with the signal coupled through path\,I when arriving at $P_2$ (or $P_1$), effectively canceling or reducing the energy coupling between the feeding ports.

\tikzmath{\d1 = 7; \d2 =5;\dant=1; \omic=1;\dmic=0.2;\want=1;\unit=1;\dy=2;\dz=2;\dhmic=1.3;\dsub=0.5;\dcore=0.5*1.5/2.1;\dh=4;\ang=70;\dhoz=\d1-2*\omic-2*\dmic-2*\unit;} 

\section{Problem formulation}\label{Sec_formulation}
The antenna structure will remain unchanged during the optimization of the decoupling network.
The antenna used in this work is designed in a separate optimization step, described in Appendix\,\ref{AntOpt}.  
The setup in Fig.\,\ref{fig:prosetup} suggests that the design of the decoupling network might be accomplished by monitoring only the reflected signals through the two feeding ports.
However, in the context of the density-based topology optimization, where the design material can be lossy during the optimization, the decoupling of port\,1 and port\,2 can simply be achieved by using a lossy material in the design domain $\Omega$, which reduces the radiation efficiency and hinders the convergence to binary materials.
A possible solution to circumvent this issue is to include the maximization of the radiated energy by the antenna elements in the problem formulation.
However, monitoring the radiated energy from the antenna is impractical as it significantly increases the computational time and memory storage during the optimization, as will be discussed later. 
Instead, we replace the antenna ports with a boundary condition that includes their impulse responses, which accounts for their mutual coupling as well.
This choice is essential for two reasons: 1) it reduces the computational cost needed for repeatedly simulating the antennas during the optimization, and 2) it facilitates the computations of gradient components associated with maximization of the radiated energy from the ports connected to the antennas.

The impulse response of an antenna depends on its geometric and material properties as well as its surrounding environment. 
As the impulse response plays a crucial role in the optimization formulation, its numerical estimation is briefly outlined here.
A full impulse response involves excitation with a Dirac pulse covering all possible frequencies.
However, antennas are typically operated within a finite frequency range, and their response outside this range is often irrelevant for modeling purposes.
Hence, in this work, we focus on estimating the \emph{band limited} impulse response by inverse Fourier transform the scattering parameters of the antenna.

The reflection coefficient $\hat\Gamma$ of a single antenna operating at frequency $f$ can be numerically calculated using 
\begin{equation}
\hat\Gamma({f})=\frac{\hat W_{\text{out}}(f)}{\hat W_{\text{in}}(f)},
\label{eq:antRef}
\end{equation}
where $\hat W_\text{in}$ and $\hat W_{\text{out}}$ are, respectively, the complex-valued incident and reflected signal at the excitation port.
In time domain, the antenna can be replaced by its impulse response $\gamma({t})$ satisfying,
\begin{equation}
W_{\text{out}}(t)=W_{\text{in}}(t)*\gamma({t}),
\label{eq:ibc_time}
\end{equation}
where $*$ is the convolution operator and the impulse response $\gamma({t})$ can be obtained from the inverse Fourier transform of $\hat\Gamma({f})$.
In this work, we utilize our FDTD code and employ a truncated time-domain \emph{sinc} function as the excitation signal $W_{\text{in}}$, whose nonzero Fourier components are used in expression\,\eqref{eq:antRef}. 
In frequency domain, the amplitude spectrum of this \emph{sinc}  function decays significantly outside the desired frequency range.
To prevent floating-point arithmetic issues arising from division by small values in equation~\eqref{eq:antRef}, a small positive regularization term is added to the denominator.

Assuming sequences of length $N$ with sampling interval $\Delta t$, the reflected signal at port\,$i$ and at time instant $t_n$, $n=0,\dots, N-1$, can be calculated using the discrete convolution
\begin{equation}
\begin{aligned}
        W_{i,\text{out}}(t_n)&=\sum_{k=0}^{n}W_{i,\text{in}}(t_k)\gamma_{i}({t_n-t_k})\Delta t,
\end{aligned}
    \label{eq:ibc_conv}
\end{equation}
where we have utilized causality of the setup.
Moreover, we may assume that $\gamma_{i}(t_n-t_k)=0$ at $k=n$, due to the presence of microstrip line segments connected to the matched feeding ports.

\begin{figure}[!htb]
\centering
\includegraphics[trim = 0mm 0mm 0mm 0mm, clip ,width=0.8\columnwidth,draft=false]{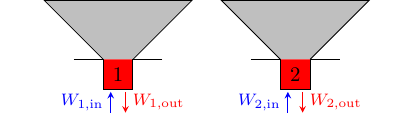}
\caption{A two-port antenna system.}
\label{fig:twoport}
\end{figure}

This technique can also be extended for the two-port antenna array shown in Fig.\,\ref{fig:twoport}, for which the system scattering matrix can be expressed as
\begin{equation}
    \begin{bmatrix}
        \hat{W}_{1,\text{out}}(f)
        \\\hat{W}_{2,\text{out}}(f)
    \end{bmatrix}=\begin{bmatrix}
        \hat{\Gamma}_{11}(f)& \hat{\Gamma}_{12}(f)
        \\\hat{\Gamma}_{21} (f)&\hat{\Gamma}_{22}(f)
    \end{bmatrix}\begin{bmatrix}
        \hat{W}_{1,\text{in}}(f)
        \\\hat{W}_{2,\text{in}}(f)
    \end{bmatrix},
\end{equation}
or in a summation form, for $i=1$, 2,
\begin{equation}
    \hat{W}_{i,\text{out}}(f)=\sum_{j=1}^{2}\hat{W}_{j,\text{in}}(f)\hat{\Gamma}_{ij}(f),
\end{equation}
where $\hat{\Gamma}_{ij}$ represents the reflection (if $i=j$) or coupling (if $i\neq j$) coefficients. 
Similarly, the time-domain impulse response relation is expressed as
\begin{equation}
    W_{i,\text{out}}(t_n)=\sum_{k=0}^{n}\sum_{j=1}^{2}W_{j,\text{in}}(t_k)\gamma_{ij}(t_n-t_k)\Delta t,
\end{equation}
in which, by causality,  the reflected signals depend solely on the signals at previous time steps, allowing the replacement of any antenna with an impulse response boundary condition to reduce the computational burden associated with full-wave simulations of the antennas.

\begin{figure}
\centering
\includegraphics[trim = 0mm 0mm 0mm 0mm, clip ,width=0.9\columnwidth,draft=false]{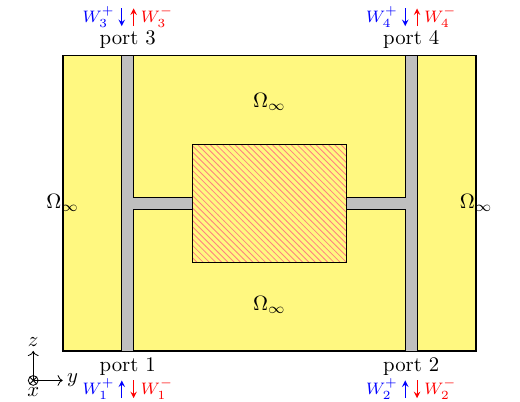}
\caption{Decoupling network with antennas at port\,3 and port\,4 are replaced by their impulse response as boundary conditions. }
\label{fig:setup_convolution}
\end{figure}
By replacing the antennas in Fig.\,\ref{fig:prosetup} with their impulse responses, the decoupling network can be simplified to a four-port network as illustrated in Fig.\,\ref{fig:setup_convolution},
where $W_i^+$ and $W_i^-$ represent the signals going into and out of the network, respectively. It should be noted that $W_3^-$ (or $W_4^-$) and $W_3^+$ (or $W_4^+$) stand for the signals traveling into the antenna and the signal reflected into the decoupling network, respectively. 
The incoming signals at port\,3 and port\,4 at time instant $t_n$ can be calculated as
\begin{subequations}
\label{DecouplingConv}
\begin{align}
       & W^+_{3}(t_n) = \notag\\
      &\quad  \Delta t\sum_{k=0}^{n}\! W^-_{3}(t_k)\gamma_{33}({t_n-t_k})+W^-_{4}(t_k)\gamma_{34}({t_n-t_k}) \\
    & W^+_{4}(t_n) = \notag\\
    &\quad\Delta t\sum_{k=0}^{n} \!W^-_{4}(t_k)\gamma_{44}({t_n-t_k})+W^-_{3}(t_k)\gamma_{43}({t_n-t_k})
\end{align}
\end{subequations}
Note that for this 4-port network, port\,3 and port\,4 only act passively and the excitation signals are only imposed into port\,1 or port\,2.
We emphasize that the convolutions \eqref{DecouplingConv} account for the mutual coupling between port\,3 and port\,4, that is, between Ant\,1 and Ant\,2 in Fig.\,\ref{fig:prosetup}.
By this treatment, the outgoing signals to the antennas ($W_3^-$ and $W_4^-$) can be considered in the formulation of the optimization problem to ensure a high radiation efficiency of the entire system, 
{as will be discussed further in Section \ref{sec:opt}.
For arrays with more antenna elements, the system's scattering matrix must be estimated, and the convolutions \eqref{DecouplingConv} can be extended accordingly.
However, it is important to note that mutual coupling between non-neighboring antennas typically diminishes with increasing distance, and the contributions to the convolutions \eqref{DecouplingConv} are primarily affected by nearby elements. 
}
 
\section{Optimization problem}\label{sec:opt}
For the electromagnetic analysis, we numerically solve the three-dimensional (3D) Maxwell's equations inside the analysis domain,
\begin{subequations}
    \begin{align}
&\frac{\partial}{\partial t} \mu \mathbf{H}+\nabla\times \mathbf{E}=\mathbf{0},
\\
&\frac{\partial}{\partial t} \epsilon \mathbf{E}+\sigma\mathbf{E}-\nabla\times \mathbf{H}=\mathbf{0},
\end{align}
\end{subequations}
where $\mu$, $\epsilon$, and $\sigma$ are the permeability, permittivity, and conductivity of the material, respectively, and $\mathbf{E}$ and $\mathbf{H}$ are the vectorial electric and magnetic fields, respectively. 
Regarding the port conditions, recall the one-dimensional (1D) model of transmission lines supporting TEM waves~\cite{hassan2014topology},
\begin{subequations}\label{Transport1D}
\begin{align}
&\frac{\partial }{\partial t}(V\pm Z_cI)\pm c\frac{\partial }{\partial z}(V\pm Z_cI)=0 \quad\text{for $z>z_0$},     \label{Transport1Da}
\\
&V+Z_cI=g(t) \quad\text{at  $z=z_0$},\label{Transport1Db}
\end{align}
\end{subequations}
where $V$ is the voltage difference, $I$ the current, $Z_c$ the characteristic impedance of the transmission line, $c$ the wave's propagation velocity, $z_0$ the interface to the transmission line, and $g(t)$ the excitation signal. 
The signs of $V$ and $I$ are defined such that the terms $V+Z_cI$ and $V-Z_cI$ represent the signals traveling in the positive and the negative direction with respect to a given direction, indicated by increasing $z$, of the transmission line. 

In this work, the antennas and the decoupling network are connected to the corresponding ports using microstrip lines, which only approximately satisfy equations~\eqref{Transport1Da}. 
We thus use full-wave modeling, and not equation\,\eqref{Transport1D}, for the microstrip lines.
However, we calculate from the field quantities equivalent voltage and current quantities at the ports.
The splitting $V\pm Z_c I$ is then used to impose the excitation signal as a boundary condition as well as to calculate the reflected signal at the port.
Although microstrip lines only support quasi-TEM modes~\cite{Pozar21microwave}, our numerical calculations show that boundary condition \eqref{Transport1Db} can be used to impose incoming signals (i.e., $W_i^+$ for $i=1,2$ in Fig.\,\ref{fig:setup_convolution}) to the microstrip lines with reflection coefficients at the ports' interfaces as low as $-22$\,dB in the frequency band of interest.
At the interface of port $i$, we use the signal $V- Z_cI$ to record the outgoing signals, that is $W_i^-$ for $i=1,2$ in Fig\,\ref{fig:setup_convolution}.

The outgoing energy at port $i$ can be evaluated as
\begin{align}\label{EnergyObjectve}
&\mathcal{E}_{i}^-=\frac{1}{4Z_c}\int_0^T(W_i^-)^2\,dt,
\end{align}
where $T$ is the total simulation time. 
The outgoing energies at the four ports of the decoupling network will be used as measures to formulate the optimization problem.
The decoupling network shown in Fig.\,\ref{fig:setup_convolution}, with the impulse response boundary conditions incorporated, satisfies the energy balance
\begin{equation}\label{EnergyBalance}
    \sum_{i=1}^4 \mathcal{E}_i^+ = \mathcal{E}_{\text{out},\Omega_\infty}+\mathcal{E}_{\Omega_\infty}+\sum_{i=1}^4 \mathcal{E}_i^- ,
\end{equation}
where \added{$\mathcal{E}_i^+$ denotes the incoming energy at port $i$}, $\mathcal{E}_{\Omega_\infty}$ is the energy loss inside the decoupling network, and $\mathcal{E}_{\text{out},\Omega_\infty}$ denotes the radiation leakage by the network, which both are assumed negligible at the frequencies of operation. 

A straightforward \added{strategy} to reduce the coupled energy between port\,1 and port\,2 is to minimize the energy $\mathcal{E}_1^-$ and $\mathcal{E}_2^-$ going back into port\,1 and port\,2.
\added{However, density-based topology optimization requires gradient-based algorithms, for which the conductivity is allowed to vary \textit{continuously} between a low and a high value.
Therefore,  a minimization involving solely $\mathcal{E}_1^-$ and $\mathcal{E}_2^-$} can be realized by introducing lossy materials that dissipate the energy inside the design domain, which will severely impact the overall radiation efficiency. 
We therefore include also the maximization of the outgoing energy $\mathcal{E}_3^-$ and $\mathcal{E}_4^-$ to the objective function, which will enforce the design material to be less lossy, that is, to be a good dielectric or a good conductor.
\added{We emphasize that} the inclusion of these two terms forces more energy to be delivered to the antennas and radiated into free space, which is consistent with the aim of reducing the coupling between port\,1 and port\,2 as well.

Thus, to minimize the mutual coupling between the feeding ports and maximize the radiated energy, we formulate the optimization problem
\begin{equation}
\begin{aligned}
 \min_{\sigma}
  F(\sigma),
  \text{ where }F(\sigma)= \log{\left( \frac{\mathcal{D}_1\left(\sigma\right) \,\mathcal{D}_2\left(\sigma\right)}{\mathcal{D}_3\left(\sigma\right)\, \mathcal{D}_4\left(\sigma\right)}\right)},
\end{aligned}
\label{eq:obj_fnopt}
\end{equation}
subject to the governing equations \added{and the imposed excitation}, with $\sigma\in [\sigma_\text{min},\sigma_\text{max}]$ denoting the electric conductivity in the design domain, where $\sigma_\text{min}$ and $\sigma_\text{max}$ represent the conductivities of a good dielectric and a good conductor, respectively.
Here we use $\sigma_{\text{min}}=10^{-4}$\,S/m and $\sigma_\text{max}=10^5$\,S/m.
The quantity $\mathcal{D}_i=(a_i+\mathcal{E}_i^{-})^{q_i}$ denotes the weighted and regularized outgoing energy at port $i$, where the parameters $a_i$ and $q_i$ can be used to control the relative scaling and emphasis of the sub-objectives. 
Based on preliminary numerical investigations, all $a_i$ and $q_i$ were set to 1, except for $q_{2}$, which was set to 0, as this choice was found to help reduce grayness in the
final designs.
Note that the terms in the denominator of problem\,\eqref{eq:obj_fnopt} are maximized, which, based on energy balance\,\eqref{EnergyBalance}, implies the minimization of $\mathcal{D}_2$.
Based on the above formulation, a decoupling structure can be designed regardless of the antenna structure, and this work can be extended to multiple antenna systems.

Solving problem~\eqref{eq:obj_fnopt} using gradient-based optimization algorithms requires calculating the derivatives of the outgoing energy at each port (that is, $\mathcal{E}_i^-$ for $i=1,2,3,4$) with respect to the conductivity changes inside the design domain. 
The first-order variation of the outgoing energy at port $i$ based on a conductivity perturbation $\delta\sigma$ in the design domain is
\begin{subequations}
    \begin{align}
\delta \mathcal{E}_{\text{i}}^-(\sigma,\delta\sigma)&= \frac{1}{2Z_c}\int_0^T W_i^-\,\delta W_i^-\,dt \\
& = \frac{1}{2Z_c}\int_0^T(V-Z_cI)(\delta V-Z_c\,\delta I)\,dt,
\end{align}
\end{subequations}
where $\delta V$ and $\delta I$ are the first-order variations of the potential difference and current at port\,$i$, respectively. 
An explicit relation between $\delta \mathcal{E}_{\text{i}}^-$ and $\delta \sigma$ can be obtained by the adjoint field method \cite{hassan2014topology} as
\begin{equation}
    \delta \mathcal{E}_{\text{i}}^-(\sigma,\delta\sigma)=-\int_\Omega\int_0^T \mathbf{E}(T-t)\cdot \mathbf{E}^*(t)\,\delta\sigma\, dt\,  d\Omega,
    \label{eq:portenergy}
\end{equation}
where $\mathbf{E}$ is the electric field in the design domain, and $\mathbf{E}^*$ is an adjoint field obtained by solving the so-called adjoint field problem, which consists of  one additional solution in the analysis domain where the recorded outgoing signals $W_{i}^{-}$, for $i=1,2,3,4$, are reversed in time and used as sources to feed their corresponding ports in the adjoint problem.

\subsection{Discretization}
The time-domain Maxwell's equations are solved using the FDTD method with the convolutional perfectly matched layer (CPML) to absorb the outgoing waves~\cite{cpml}.
The computational domain is uniformly discretized into $N_x\times N_y \times N_z$ cubic cells with an additional ten cells in each direction allocated for the CPML. 
Based on Yee's scheme, the electric field and corresponding design variables are located at each edge center\,\cite{taflove2005computational}.
Using the field data recorded during the FDTD simulations, the derivative of the port energy $W^-_i$ with respect to the conductivity $\sigma_e$ at any edge $e$ in the mesh is given by~\cite{hassan2014topology}
\begin{equation}\label{eq:DW-dsigmae}
\frac{\partial W_i^-}{\partial \sigma_{e}}=-({h)}^3\sum_{n=1}^N E^{N-n}_e
\frac{E^{*n-\frac{1}{2}}_e+E^{*n+\frac{1}{2}}_e}{2}\Delta t,
\end{equation} 
where $h$ and $\Delta t$ are the spatial and temporal discretization steps, respectively, $E_{e}$ is the discrete scalar electric field calculated by the FDTD method at edge $e$, $E^*_{e}$ is the discrete adjoint field obtained by solving the adjoint system, and $N$ is the number of total time steps.
Note the computational advantage of the adjoint field method: the derivative with respect to \textit{all} edges can be computed with only \textit{two} solutions (forward + adjoint) of the governing equations.

\subsection{Filtering}
In density-based topology optimization, the design variables are allowed to vary continuously, which leads to conductivities that can have any value between~$\sigma_\text{min}$ and~$\sigma_\text{max}$ during the optimization. 
The intermediate conductivity values are associated with high dissipative energy losses~\cite{hassan2018topology} due to ohmic losses in the design domain.
According to energy balance~\eqref{EnergyBalance}, the maximization of the outgoing energies at the antenna ports implies a minimization of the energy losses in the design domain. 
Therefore, the factors in the denominator of problem\,\eqref{eq:obj_fnopt} ensure a \textit{self-penalization} of the design conductivities toward $\sigma_\text{min}$ or $\sigma_\text{max}$, that is, towards lossless designs consisting of good dielectrics or good conductors.
The self-penalization is useful at the end of the optimization; however, it may result in a quick, premature convergence associated with a poor-performing design~\cite{Hassan14patch}. 
Filtering the design variables and using a continuation approach over the filter parameters are well-established approaches to solving this problem~\cite{hassan2014topology,bokhari2023topology,emad2020waveguide}. 

The \textit{design variables}, that is, the decision variables actually updated by the optimization algorithm, are given by a vector~$\bm{p}$, where~$p_i \in [0,1]$, with ${p}_i=1$ and ${p}_i=0$ denoting presence and absence of design material, respectively. 
The design variables are filtered by applying a weighted average in terms of a matrix~$\mathbf{A}$, where the average is performed over a neighborhood with a radius~$R$, 
\begin{equation}
    \tilde{\bm{p}}=\mathbf{A}_R \bm{p}.
\end{equation}
For the design variables near the boundary of the design domain $\Omega$, complementary variables extended outside the design domain can be used in the filtering step~\cite{clausen17OnFilter}.
In this work, the extended domain is assumed to contain vanishing complementary variables for antenna optimization, and mirror symmetry is utilized for optimizing the decoupling network. 
The design variable is then mapped to the physical conductivity using
\begin{equation}
    \sigma{(\bm{r})}=10^{9\tilde{p}(\bm{r})-4}.
\end{equation}
To reduce the losses inside the design domain, the radius $R$ of the filter is, at regular intervals during the optimization iterations, reduced using $R_{k+1}=\beta R_{k}$ with $0<\beta<1$.

Classically, matrix $\mathbf A_R$ simply encodes a local weighted arithmetic mean, that is, a local blurring of the design variables.
However, this type of linear filter may produce intermediate densities and small geometrical features, which cause difficulties in manufacturing and uncertainties in performance.
In recent years, \textit{nonlinear} filters have been proposed to mediate this problem and to achieve minimum-size control through the use of consecutive filter operators~\cite{HaWa16,hassan2018topology} over the design variable~$\bm{p}$,
\begin{equation}
    \tilde{\bm{p}}=\mathbf{F}^{K}\left(  \mathbf{F}^{K-1}\left(\dots\mathbf{F}^1\left(\bm{p}\right) \right) \right).
\end{equation}
where each $\mathbf F^k$ is a nonlinear filter operator constructed from the \textit{erode} and \textit{dilate} functions to shrink and expand the features to achieve various goals. 
By combining these two functions in a specific sequence, so-called \textit{open} and \textit{close} filters are constructed. 
The open filter fills small holes, while the close filter removes small and isolated features, both maintaining the larger-scale shape of the structure.
The combined \textit{open--close} filter is used for the antenna optimization, whereas only the open filter is used for the feeding network optimization. 
For large values of nonlinearity parameter $\alpha$, all these filters approximate a linear blurring, whereas for $\alpha\to0$, the filters increasingly well approximate the erode, dilate, open, and close operators.

The optimization procedure using nonlinear filters includes two stages and starts with a relatively large radius $R_0$ and nonlinearity variable $\alpha_0$ to avoid fast convergence to bad local minima.
In the first stage, the filter radius is gradually reduced until it reaches a finite minimum value $R_\text{min}$; however, intermediate design values (lossy design) may exist. 
In the second stage, the filter radius is fixed at $R_\text{min}$, and the nonlinearity parameter $\alpha_0$ is iteratively decreased. 
As $\alpha$ decreases, the filter's nonlinearity increases, effectively removing lossy designs and resulting in sharp copper--void boundaries.
The optimization algorithm terminates when $\alpha$ drops below a small value $\alpha_\text{min}
$.

\subsection{Numerical treatment}
To solve optimization problem \eqref{eq:obj_fnopt} using gradient-based optimization methods, we need the gradient of the objective function. 
As previously mentioned, the adjoint-field method is used to compute gradients, which requires solving one additional adjoint problem per observation included in the objective function~\cite{nikolova2006sensitivity,hassan2015time}.
For the decoupling network optimization, energies at the four ports are observed. Generally, each observation requires to solve one adjoint problem with the corresponding time-reversed signals, resulting in four adjoint problems in total. However, due to the linearity of Maxwell's equations,  
the number of adjoint problems required for the four observations of each forward problem is reduced to two.

\begin{figure}[!htb]
\centering
\includegraphics[trim = 0mm 0mm 0mm 0mm, clip ,width=0.9\columnwidth,draft=false]{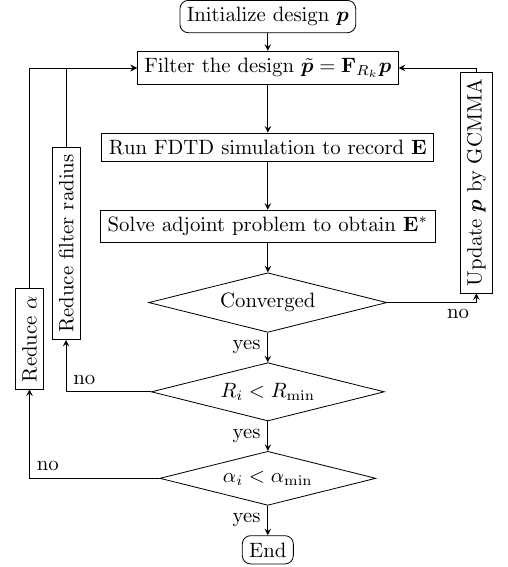}
\caption{Flowchart of the optimization algorithm. }
\label{fig:flowtopo}
\end{figure}

The optimization process starts with a uniformly distributed initial design vector $\bm{p}_0$ over the design domain $\Omega$. 
An initial radius $R_0$ and nonlinear constant $\alpha_0$ are chosen to filter the design variables. 
The FDTD method is used to solve the forward problem as well as the adjoint problem to obtain the electric field and the adjoint field for each edge inside the design domain, from which the derivatives of the objective function~\eqref{eq:obj_fnopt} can be computed, using expression~\eqref{eq:DW-dsigmae} and the chain rule.
We use the Globally Convergent Method of Moving Asymptotes (GCMMA)~\cite{SvanbergGlobally} to update the design vector.
Finally, the radius (in the first stage) and the nonlinear parameter (in the second stage) are updated until reaching their minimum values.
A flowchart of the optimization algorithm is shown in Fig.\,\ref{fig:flowtopo}. 

\section{Results}
\subsection{Simulation details}
The topology optimization part of the code is implemented in MATLAB, while the in-house 3D\,FDTD solver\,\cite{GPUFDTD}, is implemented and executed on graphics processing units (GPUs) using the CUDA toolkit.
The FDTD code is accessed from MATLAB via MEX functions. 
The code runs on one AMD zen4 node of the HPC2N cluster, which are equipped with NVIDIA H100 GPUs. 
Reduction techniques~\cite{harris2007optimizing} are employed to improve the computational efficiency of the numerical convolutions involved in the impulse-response boundary conditions.

The discretization steps for FDTD simulations are $h=0.10$\,mm in space and  $\Delta t={0.99h}/{\sqrt{3}c}$ in time. 
A~\emph{sinc} pulse with two side lobes covering a bandwidth of 0.4 GHz, centered around 2.45 GHz, is used as the excitation signal.
The number of time steps used in each FDTD simulation is $N=70\,000$ with a runtime of about 10~seconds.
The filter parameters are given in Table\,\ref{tab:paranonlinear}.
\begin{table}[!htb]
    \caption{Parameters of the nonlinear filter.}
    \centering
    \renewcommand{\arraystretch}{1.3}
\begin{tabular}{ |c|c|c|c|c|c|  }
 \hline
 Filter parameters & Initial&Min&Update equation\\
 \hline 
 {Radius $R$}&$10h$&$3h$&$R_{i+1}=0.75R_i$\\\hline 
  { Nonlinearity parameter $\alpha$}&8&$10^{-5}$&$\alpha_{i+1}=0.5\alpha_i$\\ \hline
\end{tabular}
    \label{tab:paranonlinear}
\end{table}
During each forward and adjoint simulation, the field values within the design domain $\Omega$ are recorded to compute the gradients of the objective functions. 
For each GCMMA external iteration, 0-4 inner iterations are performed to ensure convergence.
Each inner loop consists of two forward simulations: including the transmitting and receiving cases for the antenna optimization, or two single-input cases from port\,1 or port\,2 for the decoupling network optimization.

\subsection{Two-element antenna array}
\begin{figure}[!htb]
\centering
\includegraphics[trim = 0mm 0mm 0mm 0mm, clip ,width=0.9\columnwidth,draft=false]{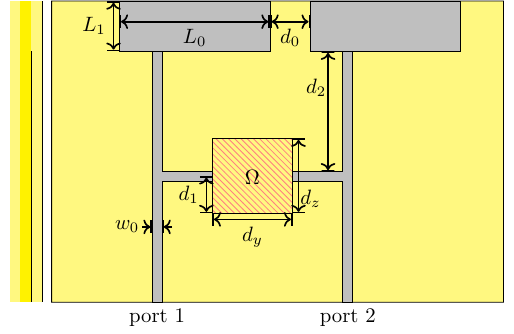}
\caption{Geometrical parameters of the two-element planar antenna array with the domain for the decoupling network included.
The system is built on a 4-layer FR-4 stackup with $\epsilon=4.5$ and 0.80\,mm thickness. 
Side (left) and top (right) views of the substrate.
The designs are placed on the top layer, and the second layer, separated by a distance $0.21$\,mm, serves as a ground plane, excluding the area beneath the antenna. $L_0=21.04$, $L_1=10.52$, $d_0=4.21$, $d_2=13.05$, $d_y=9.68$, $d_z=8.21$, $w_0=0.42$ (unit: mm). 
}
\label{fig:decouplingPara}
\end{figure}

Cross-sectional views illustrating the decoupling network, including the antenna elements and the geometrical parameters used in this work, are shown in Fig.\,\ref{fig:decouplingPara}.
The individual antenna elements are optimized to radiate around the frequency 2.45\,GHz; see Appendix\,\ref{AntOpt} for more details.
A 4-layer stackup with a total thickness of 0.8\,mm is used to build the system. 
We only utilize the top copper layer to design the antenna and the decoupling network.
The second copper layer, separated a distance 0.21\,mm from the top layer, is used as a ground plane, excluding the areas beneath the radiating patches of the antennas where copper is removed.
The material type of the three prepreg layers is FR-4 Standard TG 135--140 with 4.5 relative permittivity, which is cost-effective and well-suited for massive production.

\begin{figure}[!htb]\centering
     \subfloat[][]{
    \includegraphics[width=0.95\linewidth]{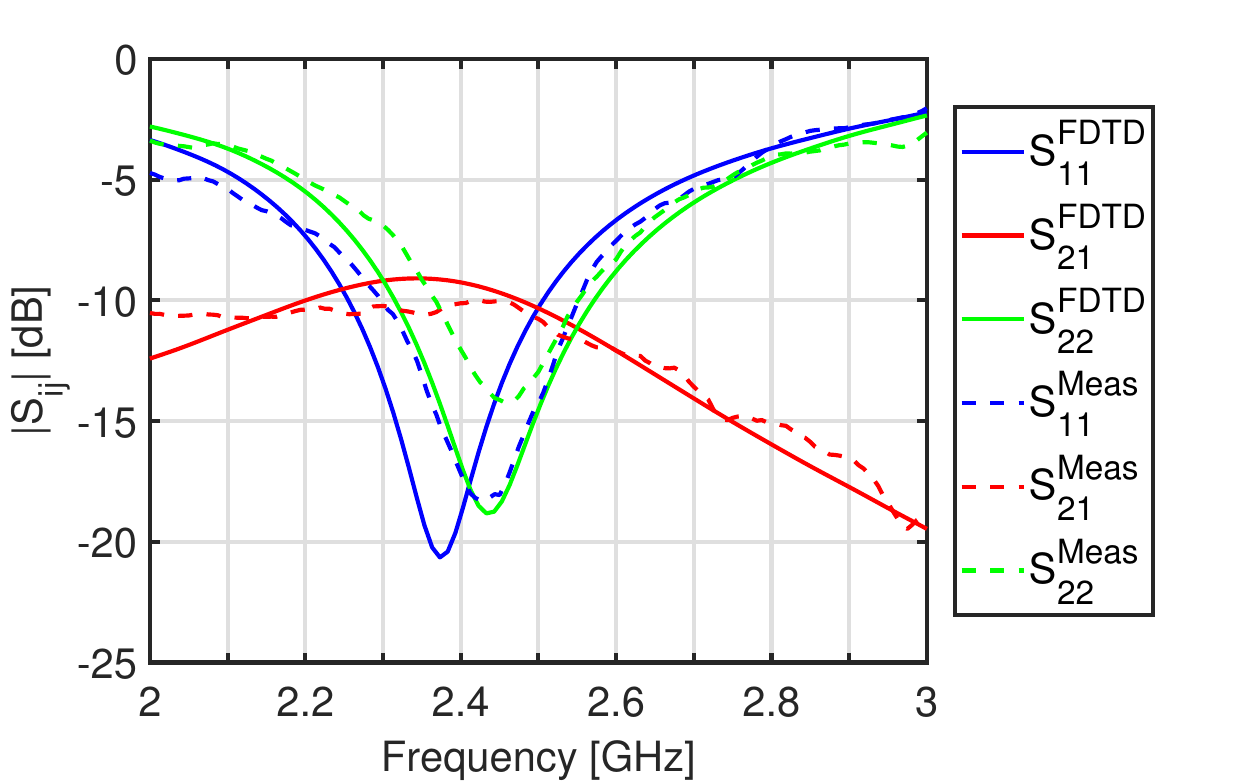}    
        \label{fig:fnorispara}}  
\vspace{5pt}        
     \subfloat[][]{
    \includegraphics[width=0.8\linewidth]{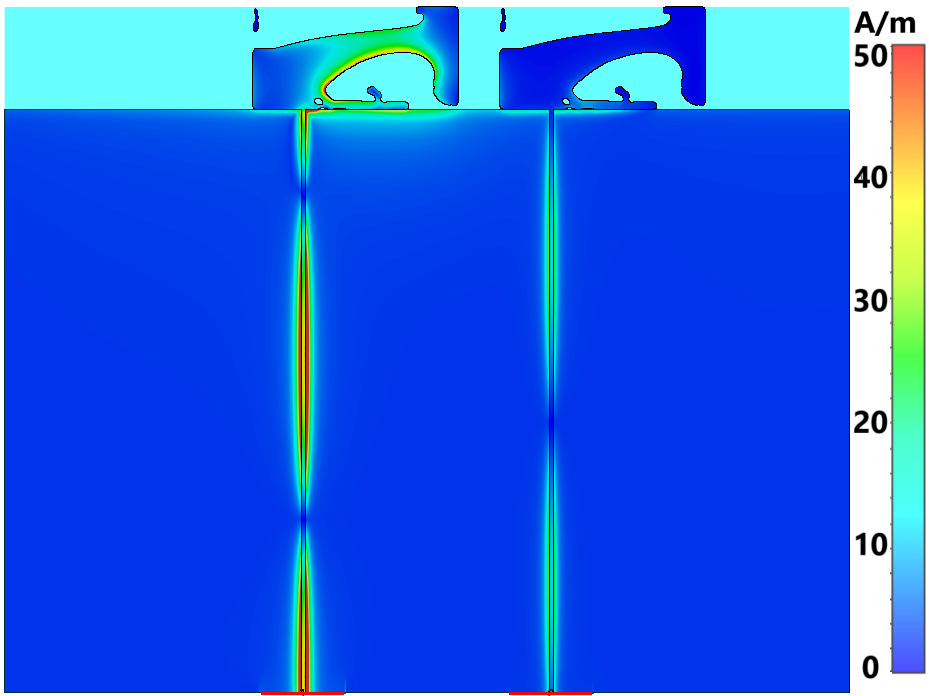}     \label{fig:fnoricst}}
\caption{(a) Simulated and measured S--parameters of the two-element antenna system without decoupling structure. (b) Current distribution of the two-element antenna system at 2.45\,GHz, simulated using the CST Studio Suite~\cite{cst}, showing a noticeable current coupled to port\,2 when port\,1 is excited.}
\end{figure}

We simulated and measured the performance of the two antennas without the decoupling network.
Fig.\,\ref{fig:fnorispara} shows the magnitude of the S--parameters of the two antennas, comparing the simulated results using our FDTD code with the measured S--parameters. 
The slight differences between the simulation and measurements could be attributed to material or geometrical uncertainties in the manufactured designs. 
The measurement is performed using a Rohde \& Schwarz ZND Vector network analyzer (VNA), and the setup is shown in Fig.\,\ref{fig:measruefig}.
Compared to the performance of a single antenna element in free space, as shown in Appendix\,\ref{AntOpt}, which exhibits $\approx-20$\,dB dip in $|S_{11}|$ around 2.5\,GHz, the near-field interaction causes the dips in the reflection coefficients $|S_{11}|$ and $|S_{22}|$ of the array elements to shift toward lower frequencies by $\approx$ 100\,MHz and 50\,MHz, respectively.
In addition, the simulated mutual coupling between the two antennas has a peak value of $|S_{21}|\approx -9$\,dB around 2.35\,GHz. 
Fig.\,\ref{fig:fnoricst} shows the current distribution of the two antennas at 2.45 GHz, simulated using the CST Studio Suite, where port\,1 is used for excitation and port\,2 is matched.
There is a noticeable current coupled to port\,2, indicating strong coupling between the two antennas.

\begin{figure}
\begin{center}
    \includegraphics[height=0.27\linewidth]{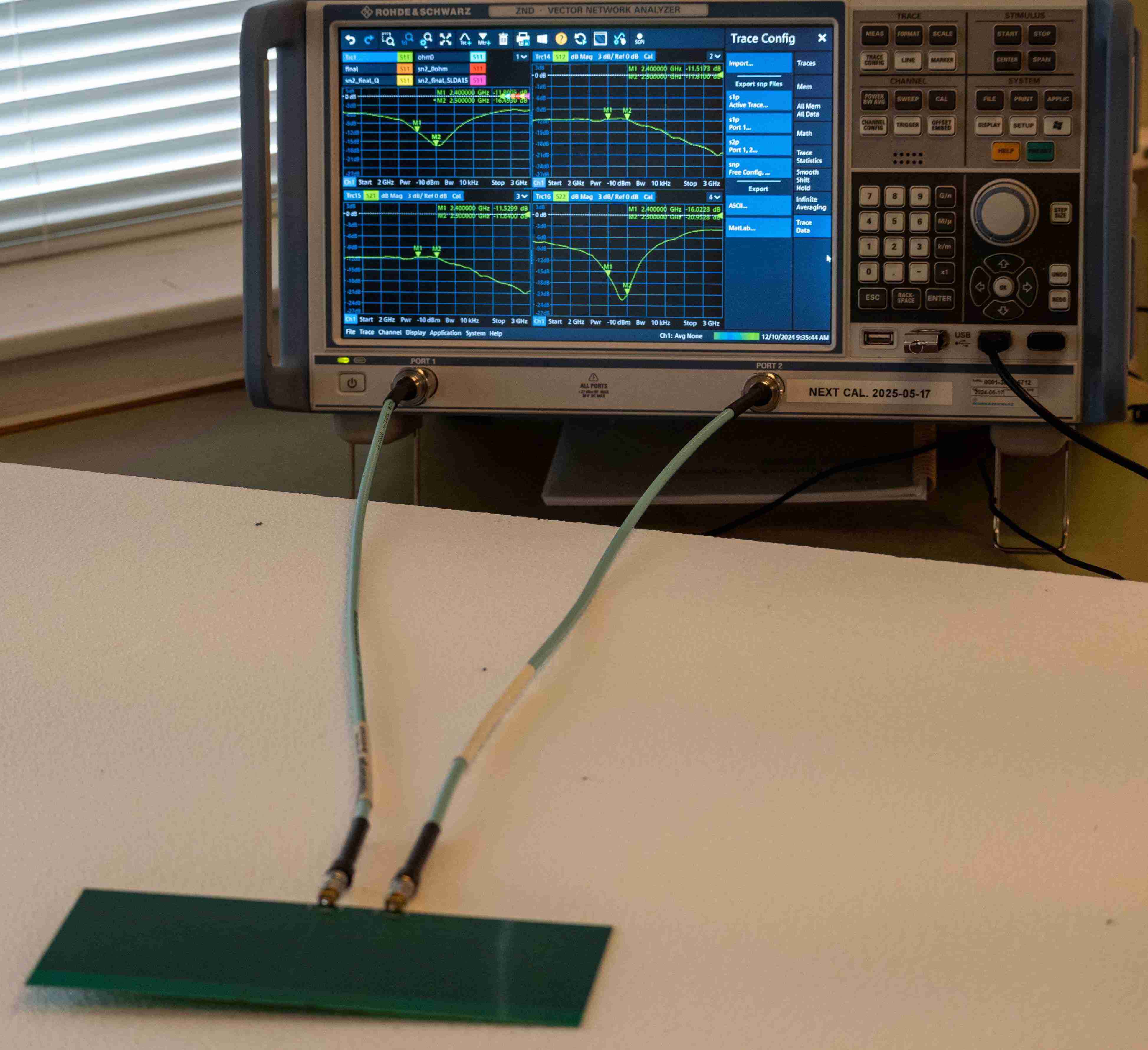}
        \includegraphics[height=0.27\linewidth]{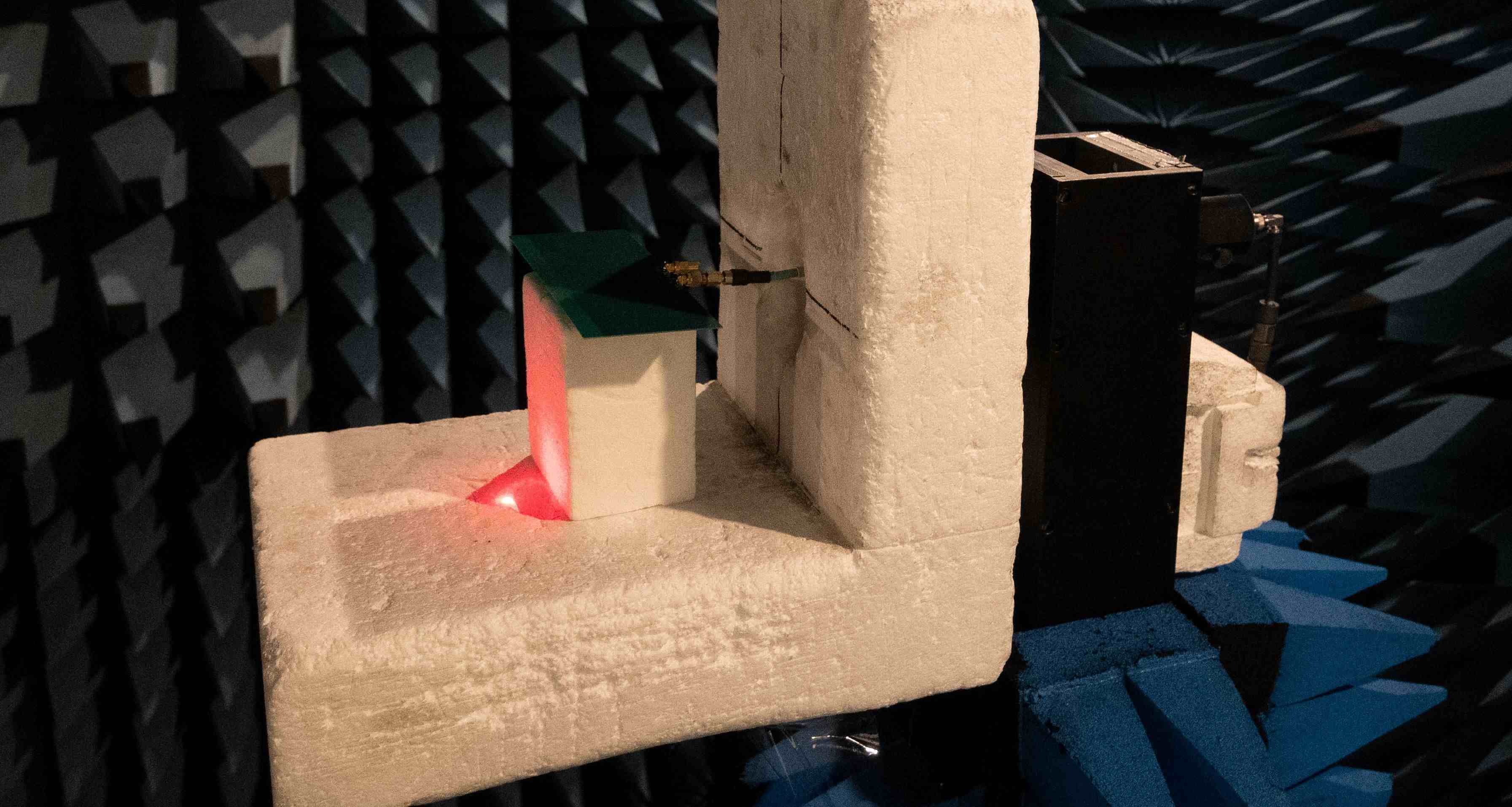}
         \includegraphics[height=0.27\linewidth]{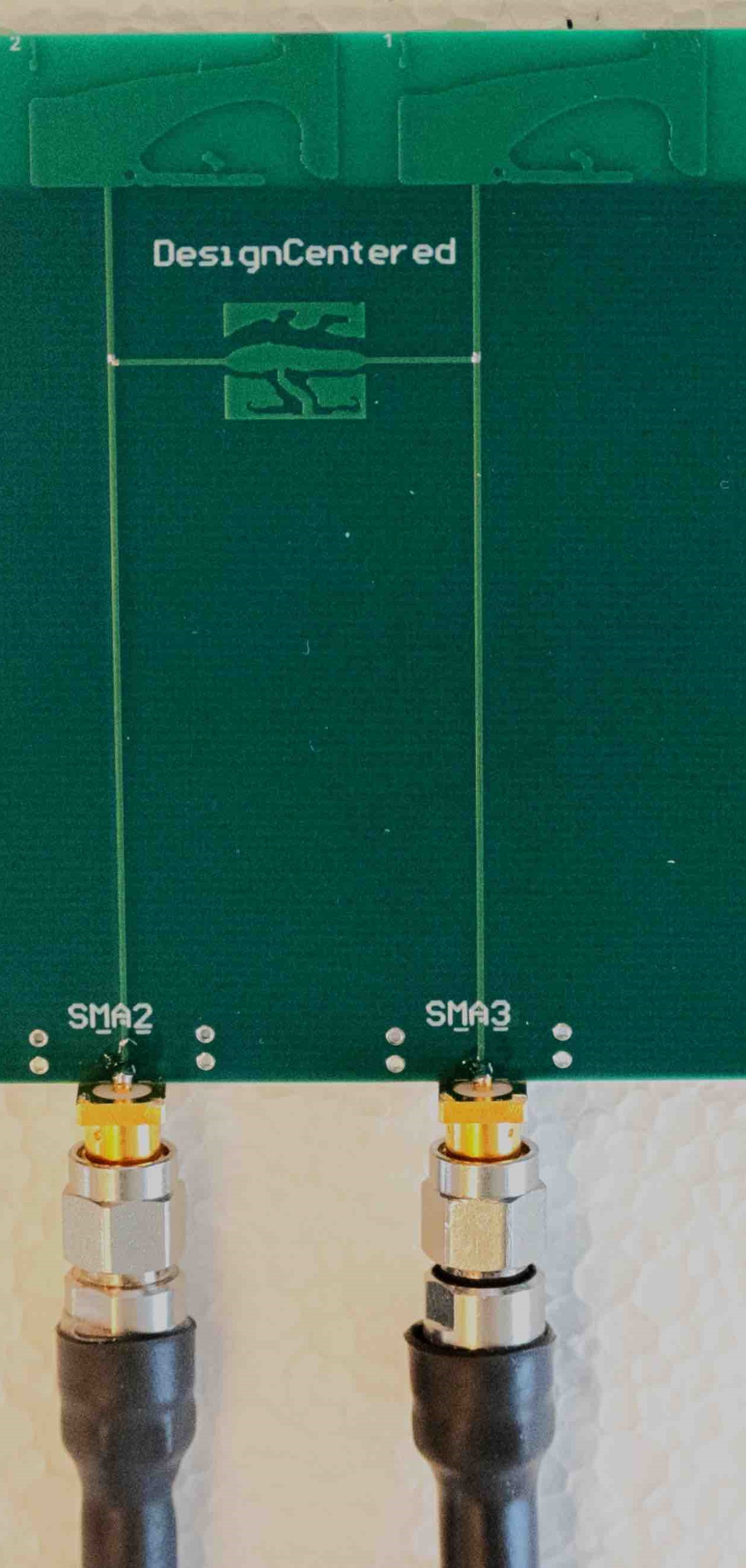}
\end{center}
    \caption{Photos of the measurement setup and a prototype of a decoupling network design.}
    \label{fig:measruefig}
\end{figure}

 Before optimizing the decoupling network, two full-wave FDTD simulations (one for each port input) are conducted to obtain the reflection and coupling coefficients, which we use for estimating the impulse responses in the frequency band of interest by the procedure discussed in Section~\ref{Sec_formulation}. 
In these simulations, the two antennas are connected to ports\,3 and port\,4 using short microstrip lines. 
For the chosen discretization, this treatment results in a reduction of the computational domain by 72\%, from $50\times480\times315$ Yee cells to $40\times272\times195$ Yee cells,
during the optimization of the decoupling network.
The computational efficiency improves significantly when this approach is used to optimize feeding networks for large antennas, since such problems are often computationally demanding.

\subsection{Decoupling network optimization}
Based on the network model given in Fig.\,\ref{fig:setup_convolution}, we solve the optimization problem \eqref{eq:obj_fnopt}.  
The design domain $\Omega$ is planar and has dimensions, as defined in Fig.\,\ref{fig:decouplingPara}, of $d_y=92h=9.68$\, mm and $d_z=78h =8.21$ mm, resulting in $29\,044$ design variables associated with the edges of Yee cells in $\Omega$. 
{The width of the decoupling structure is limited to approximately one-third of the distance between the microstrip lines to prevent interference with the feeding lines and reduce potential radiation leakage.}
The microstrip lines, feeding the domain $\Omega$ at the middle side, are separated by a distance $d_2=13.05$\,mm from the antenna patches, see Fig.\,\ref{fig:decouplingPara}.
The initial design corresponds to a uniform density vector of value $\rho_i=0.7$, except for small regions of size $R\times w_0$ at the interface between $\Omega$ and the microstrip lines, where $\rho_i=1$ is used.
These fixed regions provide a stable connection between the design domain and the feeding lines during the optimization.

\begin{figure}[!htb]   
\begin{subfigure}[t]{0.5\textwidth}
\includegraphics[trim = 0mm 0mm 0mm 0mm, clip ,width=0.95\columnwidth,draft=false]{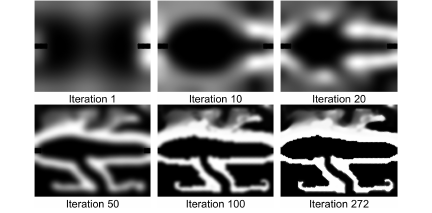}
\caption{}
\label{fig:FNoptobj3process}
\end{subfigure}  
\begin{subfigure}[t]{0.5\textwidth}
\begin{center}
\includegraphics[width=0.95\linewidth]{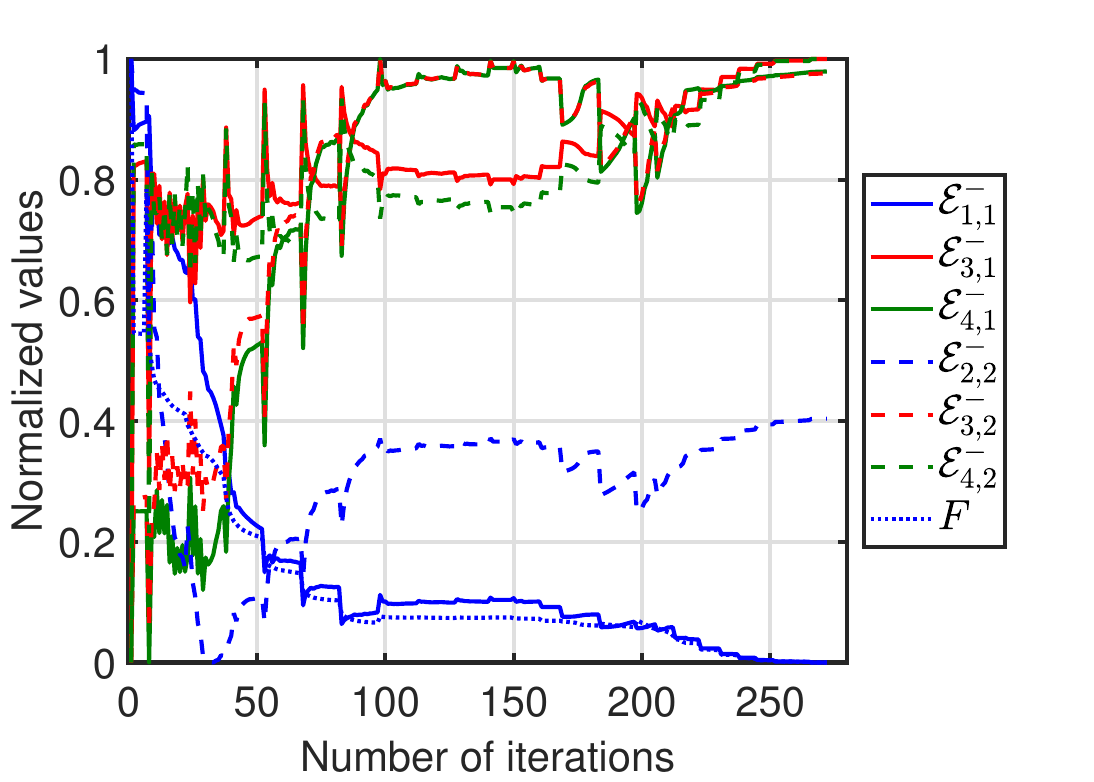}
\end{center}
\caption{}
\label{fig:fnmiddle}
\end{subfigure}
\caption{Decoupling network optimization. (a) Snapshots demonstrating the evolution of the decoupling structure. b) iteration history of the normalized objective function $F$ and the outgoing energies at the ports of the decoupling network.}  
\label{fig:History}
\end{figure}

Fig.\,\ref{fig:FNoptobj3process} shows snapshots demonstrating the development of the decoupling structure during the optimization process.
In the early iterations, the design variables are highly blurred and only large features evolve, emphasizing a need for a connection/coupling route through the design domain.
In the beginning, the structure was nearly symmetric.
After around 50 iterations, the upper part of the design domain becomes isolated from the route connecting the two stubs. 
As the iterations progress, the blurriness decreases, and small features gradually emerge.
After 272 iterations the structure is nearly binarized, except for the upper part of the design domain, which maintains some gray material, a phenomenon we discuss further below.

Fig.\,\ref{fig:fnmiddle} shows the iteration history of the normalized objective function along with its constituent terms.
The reflected energies $\mathcal{E}_{1,1}^-$ and $\mathcal{E}_{2,2}^-$ are reduced more than 60\% compared to the initial values, while the outgoing energies into the two antennas, $\mathcal{E}_{3,i}^-$ and $\mathcal{E}_{4,i}^-$ \added{with $i$ denoting the excitation port}, are generally increasing. 
Due to the gray materials at the beginning of the optimization, the outgoing energies towards the antennas, $\mathcal{E}_{4,1}^-,\mathcal{E}_{3,2}^-$, are relatively low, while the energies reflected into the excitation port, $\mathcal{E}_{1,1}^-,\mathcal{E}_{2,2}^-$, are high, suggesting that nearly all energies are consumed in the gray material or reflected back into the port, respectively. 
The intermediate values are gradually reduced as the optimization proceeds, resulting in increased energy transmission, and the evolving structure helps in reducing the energy reflection. 
The development of the individual terms in the objective is not monotonically evolving, which is due to changes in their relative contribution to the objective during the optimization process.
However, the main objective $F$ is monotonically decreasing, except at a few exceptional instances following an update of the filter parameters.

\begin{figure}[!htb]\centering
     \subfloat[][]{
    \includegraphics[width=0.95\linewidth]{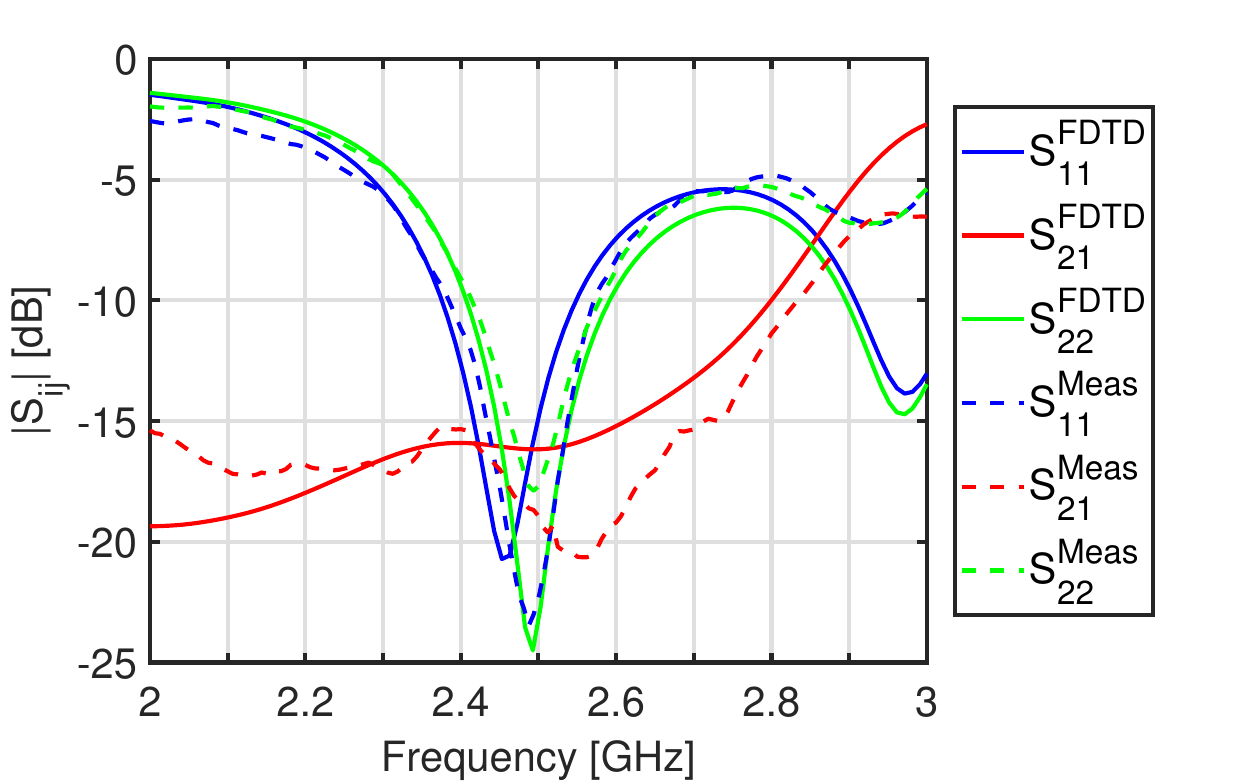}    
        \label{fig:FNobj3Spara}}  
\vspace{5pt}        
     \subfloat[][]{
    \includegraphics[width=0.8\linewidth]{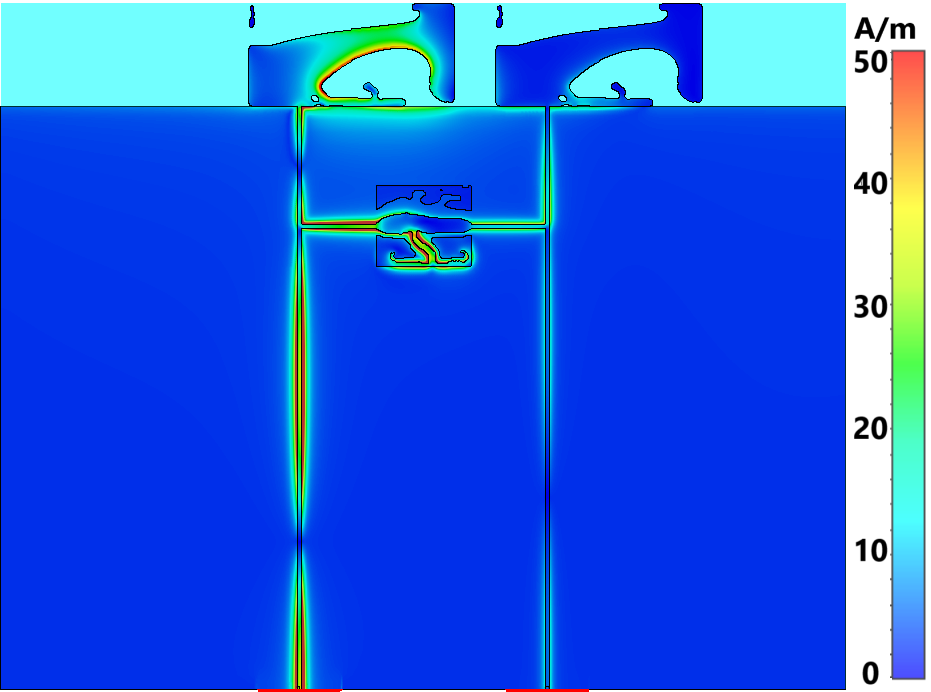}     \label{fig:fnmidcst}}
\caption{(a) Simulated and measured S--parameters of the two-element antenna system with the optimized decoupling structure. (b) Current distribution at 2.45\,GHz showing a noticeable decrease in the current coupled to port\,2 when port\,1 is excited.}
\end{figure}

The optimized design is thresholded at a density value of 0.5, of which values below 0.5 are mapped to void and values above 0.5 are mapped to copper.
We evaluate the performance of the final optimized decoupling structure, integrated with the two-element antenna array system.
Fig.\,\ref{fig:FNobj3Spara} shows the simulated and measured S--parameters of the network.
The mutual coupling has been significantly reduced to lower than $-16$\,dB, and the amplitudes of $S_{11}$ and $S_{22}$ are essentially maintained with negligible frequency shifts.
Fig.\,\ref{fig:fnmidcst} shows the surface current distribution of the decoupling network at 2.45\,GHz when port\,1 is excited.
The current distribution exhibits large values in the decoupling structure, demonstrating a strong interaction with the feeding signal. 
The high current amplitude between the decoupling structure and the second antenna indicates destructive interference between the signal resulting from the antennas' mutual coupling and the signal transmitted through the decoupling structure.

\begin{figure}[!htb]
    \subfloat[][]{  \hspace{-12pt}\includegraphics[width=0.5\linewidth]{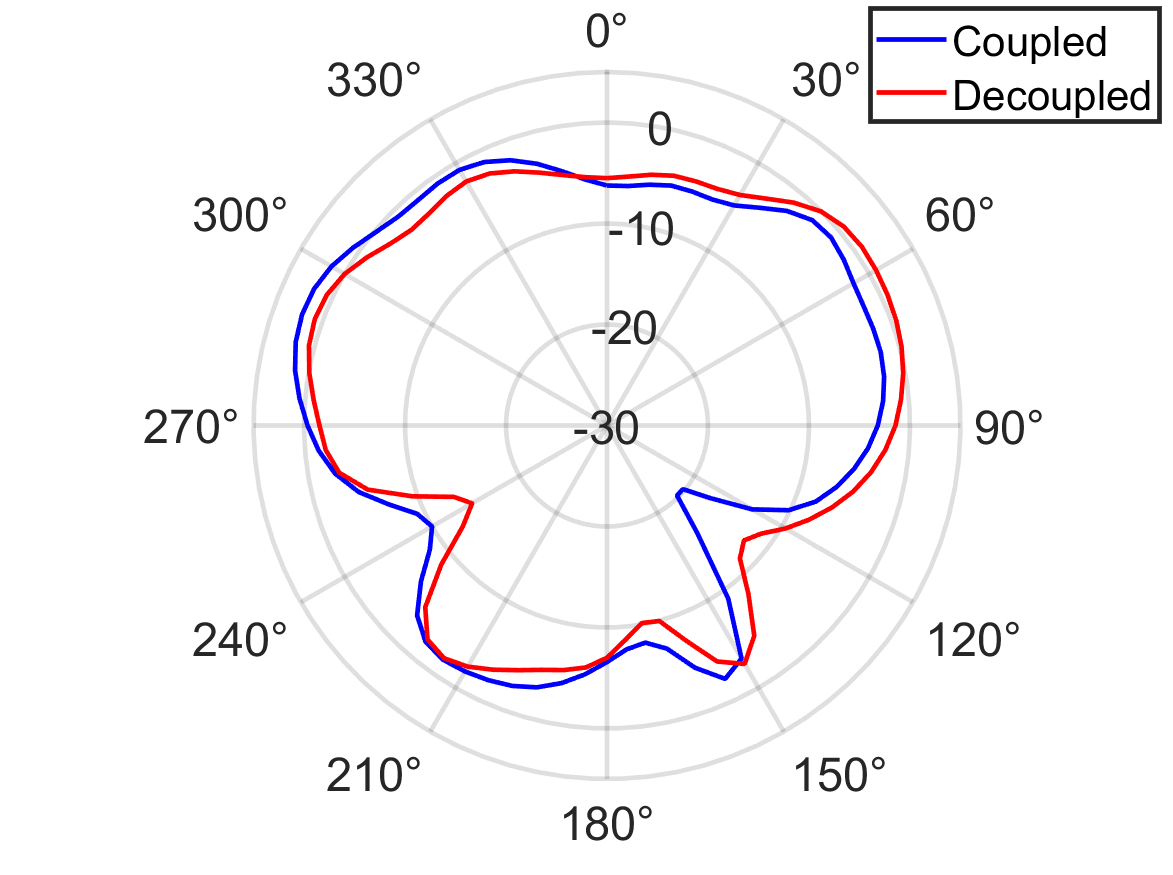}}
     \subfloat[][]{
    \hspace{-5pt}\includegraphics[width=0.5\linewidth]{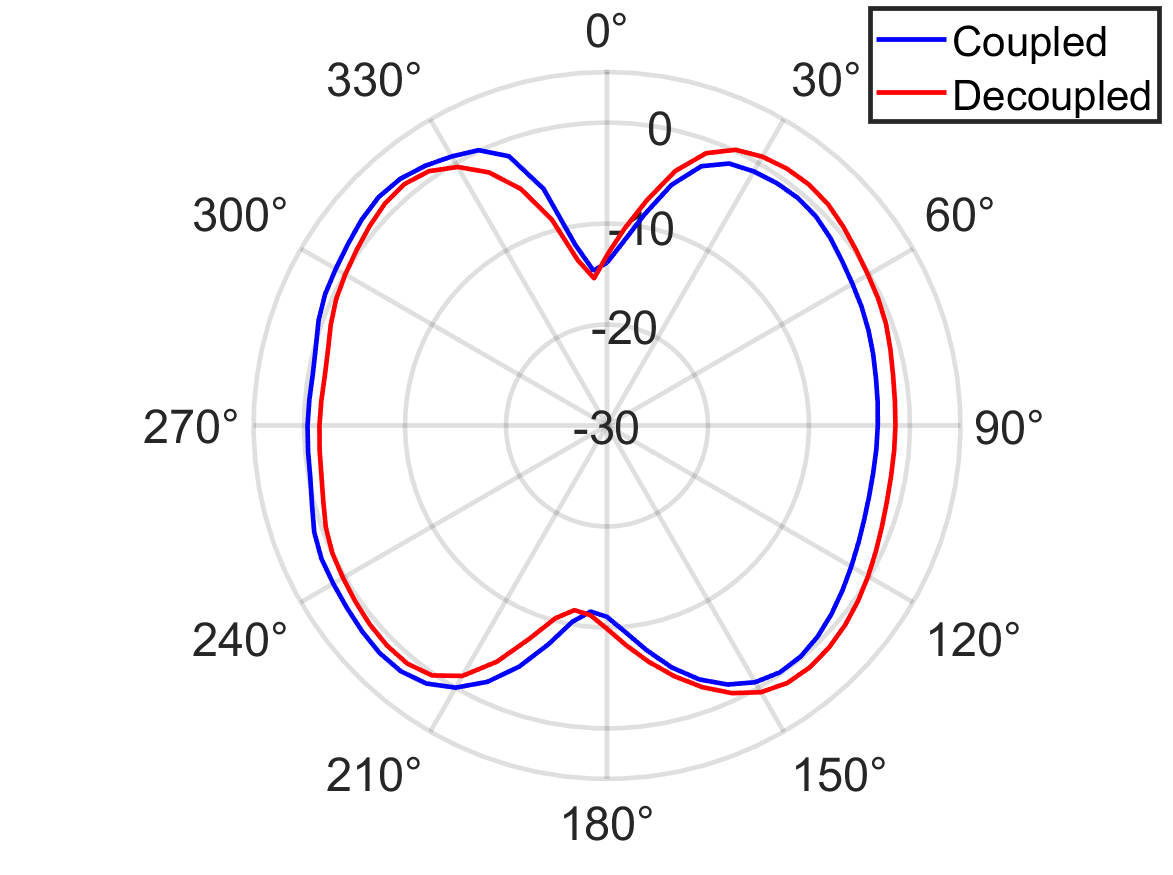}}
    \caption{Measured radiation patterns (unit: dB) of the antenna array with and without the decoupling network at 2.45\,GHz in (a) $yz$ plane and (b) $xy$ plane.}
    \label{fig:pattern}
\end{figure}

The current distribution primarily shows high values in the bottom half of the design domain, while the upper portion supports a negligible current. In other words, the upper half of the design domain has no significant influence on the functionality of the decoupling structure. 
This observation aligns with the design evolution shown in Fig.\,\ref{fig:FNoptobj3process}, where the upper area of the final design retains gray material. 
Early during the optimization, these areas become disconnected and isolated from the signal sources, making their contributions to the gradient vector negligible. 
This behavior is consistent with gradient-based optimization methods, which tend to converge to local optima. 
The far-field pattern of the antenna array is measured at 2.45 GHz, and  Fig.\,\ref{fig:pattern} compares the array with and without the decoupling network when port\,1 is excited. 
The decoupling structure has little impact on the radiation pattern, confirming its intended role as a guided structure that diverts the signals to reduce the coupling between the feeding ports.

\begin{figure}[!htb]
    \begin{subfigure}[t]{0.5\textwidth}
    \begin{center}
     \includegraphics[trim = 0mm 0mm 0mm 0mm, clip ,width=0.95\columnwidth,draft=false]{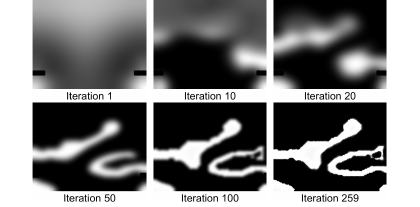}
      \end{center}
         \caption{}
         \label{fig:FNoptobj3highprocess}
    \end{subfigure}
    
   \begin{subfigure}[t]{0.5\textwidth}
   \begin{center}
\includegraphics[width=0.95\linewidth]{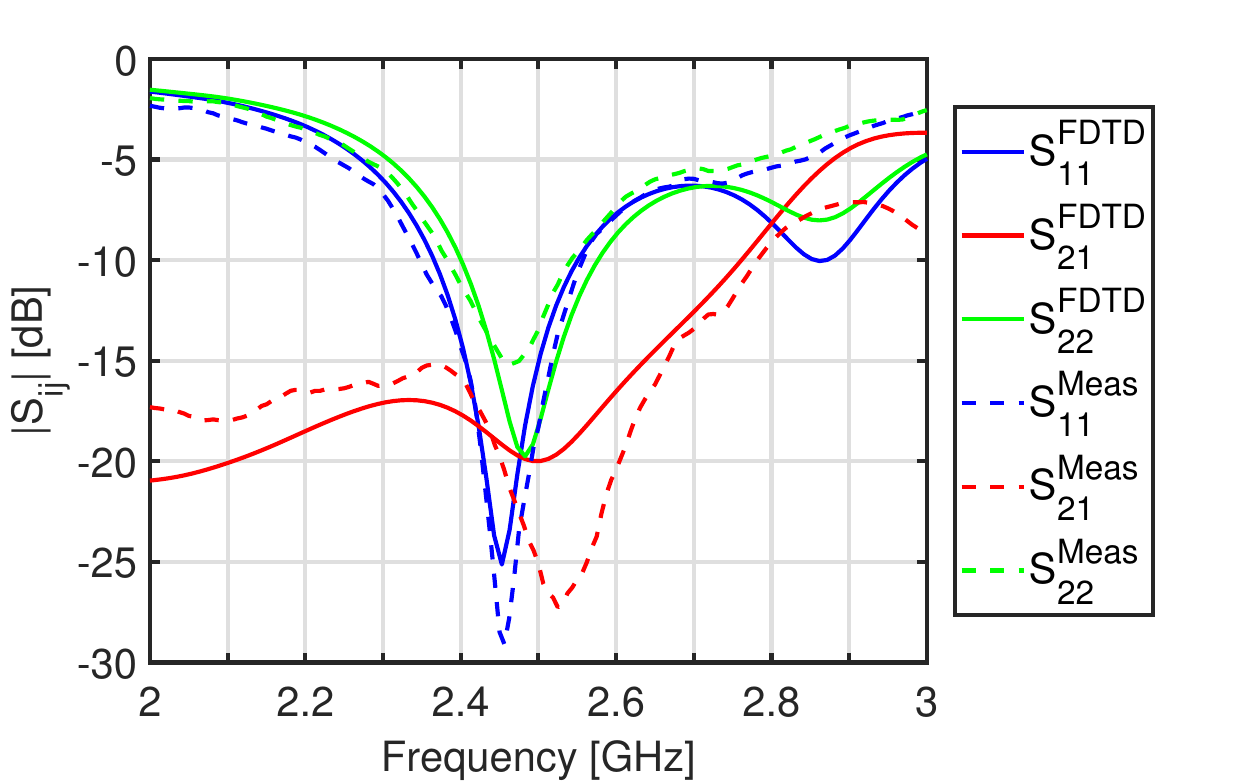}
\caption{}
    \label{fig:high25obj_spara}
\end{center}
    \end{subfigure}
    
\vspace{5pt} 
   \begin{subfigure}[t]{0.5\textwidth}
   \begin{center}
\includegraphics[width=0.8\linewidth]{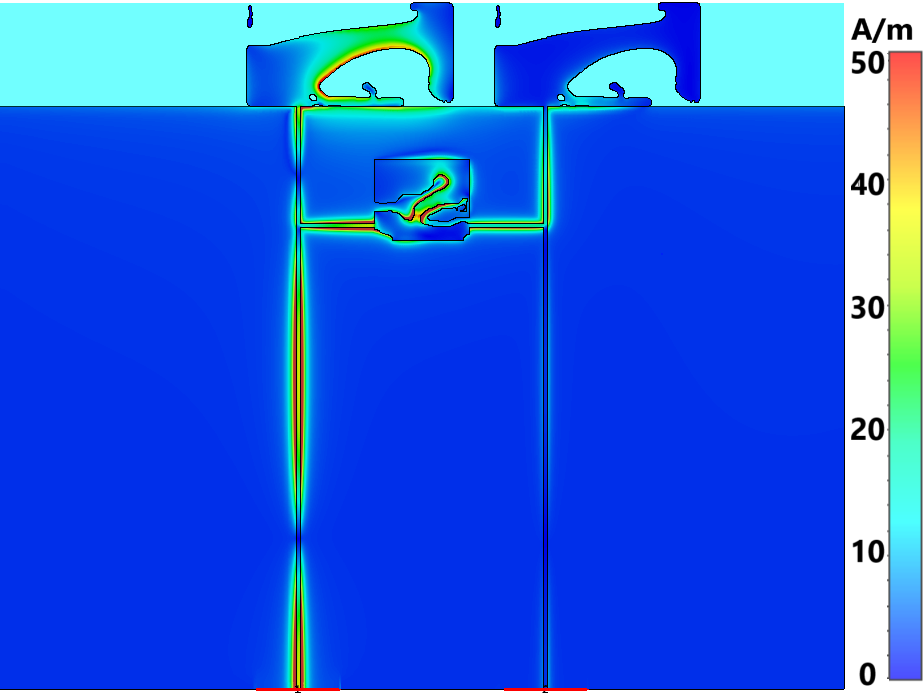}
\caption{}
\label{fig:high25}
\end{center}
    \end{subfigure}
    \caption{Decoupling Design\,II. (a) Snapshots showing the evolution of the design. (b) Simulated and measured S–parameters. (c) Current distribution at 2.45\,GHz when port\,1 is excited and port\,2 is matched.}  
    \label{fig:DesignII}
\end{figure}

Motivated by the above discussion and aiming to enhance the performance of the decoupling structure, we shift the design domain upwards by a distance of 2.63\,mm.
The new optimization results are shown in Fig.\,\ref{fig:DesignII}.
The final design, obtained after 259\,iterations, is fully connected and contains almost entirely binary variables, see Fig.\,\ref{fig:FNoptobj3highprocess}.
The simulated and measured S--parameters of Design\,II, shown in Fig.\,\ref{fig:high25obj_spara}, demonstrate a good correlation and exhibit a dip around 2.5\,GHz.
This results in more than 10\,dB ($\approx$ 15\,dB) reduction in simulated (measured) mutual coupling compared to the array without a decoupling structure.
Design\,II has a higher capacity to reduce the coupling with the fully connected structure.
The current distribution shown in Fig.\,\ref{fig:high25} illustrates the destructive interference between the two signal paths, as discussed earlier, with a larger portion of the decoupling structure actively contributing to the processing of the signals' interference.
Only a tiny current is coupled to port\,2 and the two antennas are highly decoupled.
Table \ref{tab:comparison} presents a comparison between the proposed design and several recently published works.
While a direct comparison is challenging due to differences in operating frequencies, substrate materials, and design methodologies across the referenced studies, the proposed decoupling network is among the smallest antennas and decoupling network sizes.
We emphasize that all optimized structures represent local optima of the formulated optimization problem.
Varying the optimization parameters might lead to different solutions, involving various tradeoffs among reflection, coupling, and energy transmission at the ports.

\begin{table*}
\centering
\caption{Comparison with other decoupling networks.}
    \label{tab:comparison}
\begin{tabular}{ |c|c|c|c|c|c|c|c|c|  }
 \hline
\multirow{2}{*}{Ref}& \multirow{1}{*}{Center } & \multirow{2}{*}{Method$^1$ }&\multirow{2}{*}{Type$^2$}&\multirow{2}{*}{Ant. size}&\multicolumn{2}{|c|}{$|S_{21}|$\,(dB)$^3$}&\multirow{1}{*}{Decoupling }&Edge-to-edge\\\cline{6-7}
& \multirow{1}{*}{Fre.\,(GHz)}&&&&w/o&w/&structure size$^4$&distance\\\hline
\multirow{1}{*}{\cite{li2021decoupling}}&{2.45(5.5)}&THE\&MA&DN&$0.16\lambda_0\times0.090\lambda_0$&-8(-13)& -27(-25)&$0.09\lambda_0\times0.34\lambda_0$&$0.043\lambda_0$\\\hline

\multirow{1}{*}{\cite{pei2021low}}&{3.16}&THE&PS&$0.37\lambda_0\times0.32\lambda_0$&-7&-18&$0.02\lambda_0\times0.32\lambda_0$&$0.027\lambda_0$\\\hline

\multirow{1}{*}{\cite{zhang2015line}}&{4.05}&EXP&NL&$0.20\lambda_0\times0.20\lambda_0$&-12&-30&N/A&$0.029\lambda_0$\\\hline

\multirow{2}{*}{\centering\cite{zhu2019design}}&\multirow{2}{*}{5.8}&\multirow{2}{*}{G-TO}&\multirow{2}{*}{DGS}&\multirow{2}{*}{$0.22\lambda_0\times0.22\lambda_0$}&\multirow{2}{*}{-5}&\multirow{2}{*}{-26}&TOP:$0.46\lambda_0\times0.027\lambda_0$&\multirow{2}{*}{$0.034\lambda_0$}\\
&&&&&& &BOT:$0.46\lambda_0\times0.052\lambda_0$&\\\hline

\multirow{1}{*}{\cite{ding2020novel}}&2.35&MA&DGS&$0.16\lambda_0\times0.039\lambda_0$&N/A&-25&BOT:$0.34\lambda_0\times0.078\lambda_0$&$0.078\lambda_0$\\\hline

\multirow{1}{*}{\cite{whaleswarm2025}}&2.4&MA&NL&$0.20\lambda_0\times0.096\lambda_0$&-19&-27&N/A&$0.016\lambda_0$\\\hline

\multirow{1}{*}{This work}&2.45&G-TO&DN&$0.17\lambda_0\times0.086\lambda_0$&-10&-20&$0.079\lambda_0\times0.067\lambda_0$&$0.034\lambda_0$\\\hline
\multicolumn{9}{l}{$^1$Design method based on theory (THE), experience (EXP), metaheuristic algorithms\,(MA), or gradient-based topology optimization\,(G-TO).}\\
\multicolumn{9}{l}{$^2$Decoupling techniques: DN\,(decoupling network), PS\,(phase shift), NL\,(neutralization line), DGS\,(defected ground structure).}\\
\multicolumn{9}{l}{$^3$w/o and w/ represents without and with the decoupling structure, respectively. N/A represents where data is not available.}\\
\multicolumn{9}{l}{$^4$TOP/BOT represents the decoupling structure designed on the top/ground layer. Default is TOP, if not specified.}
\end{tabular}
\end{table*}

\section{Conclusion}
This work proposes an efficient density-based topology optimization method to design decoupling networks for antenna arrays. 
Impulse responses, accounting for the mutual coupling, are employed through a time-domain boundary condition, and a gradient-based topology optimization problem is formulated to minimize the mutual coupling and maximize the radiated energy to the antenna system. 
To demonstrate the concept, decoupling networks for a two-element antenna array system operating around 2.5\,GHz are presented. 
{While the concept is presented using a two-element narrowband array, the proposed topology optimization framework is extendable to multiple frequencies and wideband systems. 
In fact, the use of a time-domain approach makes the algorithm even more suitable for targeting wideband systems, offering a more efficient computation.}
The antenna element is optimized in a separate phase; however, the concept can be applied to other types of antennas as well. 
The design configuration and the choice of optimization parameters play crucial roles in achieving high-performing designs. 
Two decoupling networks were numerically and experimentally investigated, demonstrating the effectiveness of the decoupling structures in reducing the mutual coupling and maintaining a good matching with the feeding ports. 
All the optimized designs in this work were optimized using a pixelation approach over more than 29\,000 design variables, resulting in novel designs.
The proposed approach might be applied to tackle interference-related problems, where the coupling includes not only near-field interactions but also other types of interference.

\section*{Acknowledgment}

Partial funding for this work is provided by  the Swedish strategic research program eSSENCE as well as the Swedish Research Council, grant 2018-03546. 
The computations were performed on resources provided by the High Performance Computing Center North (HPC2N). 
The authors thank Johan Haake for valuable discussions on prototyping the designs.

\section{Appendix}
\label{AntOpt}
\subsection{Antenna optimization}

\begin{figure}[!htb]
\centering
\includegraphics[trim = 0mm 0mm 0mm 0mm, clip ,width=0.9\columnwidth,draft=false]{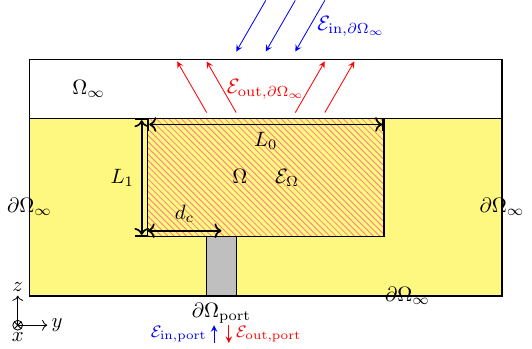}
\caption{Setup for antenna design optimization. A 50 Ohm microstrip line connects the feeding port to the design domain $\Omega=L_0\times L_1$, where the conductivity distribution of copper is to be optimized.
$L_0= 21.04$\,mm, $L_1= 10.52$\,mm, $d_c=5.26$\,mm, and the discretization step $h=0.10$\,mm, resulting in 40\,300 design variables.}
\label{fig:antdesign}
\end{figure}

In this section, we detail the design of the antenna element featured in the main text.
Fig.\,\ref{fig:antdesign} illustrates the setup for optimizing a single antenna element, where $\Omega_\infty$ is the analysis domain, $\Omega$ denotes the design domain, and the boundary  $\partial\Omega_\text{port}$ is used to impose/record signals into/from the analysis domain $\Omega_\infty$ through the microstrip line. 
This system satisfies the energy balance~\cite{hassan2014topology}
\begin{equation}\label{AntEnergyBalace}
\mathcal{E}_{\text{in,port}}+\mathcal{E}_{\text{in},\partial\Omega_\infty}=\mathcal{E}_{\text{out,port}}+\mathcal{E}_{\text{out},\partial\Omega_\infty}+\mathcal{E}_\Omega,
\end{equation}
where the total incoming energy from the port $\mathcal{E}_{\text{in},\text{port}}$ and exterior waves $\mathcal{E}_{\text{in},\Omega_\infty}$  equals the total outgoing energy $\mathcal{E}_{\text{out},\text{port}}+\mathcal{E}_{\text{out},\partial\Omega_\infty}$ and energy dissipated as ohmic loss $\mathcal{E}_\Omega$ in the analysis domain. 

Energy balance \eqref{AntEnergyBalace} can be used to formulate antenna optimization problems based on various types of excitation.
For a given incoming signal through the feeding port $\mathcal{E}_\text{in,port}$, and no exterior waves sources---that is, designing the antenna based on its transmitting mode---one may consider either to minimize the reflected energy $\mathcal{E}_\text{out,port}$ or to maximize the outgoing energy $\mathcal{E}_{\text{out},\partial\Omega_\infty}$.
The minimization of $\mathcal{E}_\text{out,port}$ would unfortunately result in a lossy design that dissipates all the energy and radiates nothing.
On the other hand, the maximization of $\mathcal{E}_{\text{out},\partial\Omega_\infty}$ requires observation of transmitted waves over a closed surface surrounding the antenna, which demands significant memory resources. 
A third alternative is to maximize the received energy $\mathcal{E}_{\text{out},\text{port}}$, given the excitation from far-field exterior sources.
In other words, the antenna is designed based on its receiving mode, relying only on the observation of the outgoing signal at the feeding port.
However, using only this last choice, the design algorithm often converges to non-satisfying designs, exhibiting a high reflection coefficient.
Therefore, we opted to combine the first and third choices, formulating the objective function as
\begin{equation}
\begin{aligned}
&\min_{\sigma} 
F(\sigma),\text{ where }F(\sigma)=\log\left(\frac{\mathcal{D}_1\left(\sigma\right)}{\mathcal{D}_2(\sigma)}\right),
\end{aligned}
\label{eq:obj_antopt}
\end{equation}
subject to the governing equations. 
Similar to problem\,\eqref{eq:obj_fnopt}, $\mathcal{D}_1$ and $\mathcal{D}_2$ represent the weighted and regularized outgoing energy at the antenna feeding port for the transmitting and receiving modes, respectively, with the selected parameters $a_1=a_2=1$, $q_1=0.2$, and $q_2=0.1$.
Being in the denominator of optimization problem \eqref{eq:obj_antopt}, the maximization of the received energy drives convergence to lossless designs, while the minimization of the reflected energy (that is, the numerator term) promotes designs with favorable reflection coefficients. 
For the receiving mode, two $y-$\, polarized plane waves are injected, including one with a propagation direction normal to the design domain (positive $x$) and the other normal to the top side of the substrate (negative $z$), respectively. 
The plane wave is imposed in FDTD simulations using the total-field scattered-field technique~\cite{taflove2005computational}. 

\begin{figure}[!htb]
\centering
\begin{subfigure}[t]{0.5\textwidth}
\includegraphics[trim = 0mm 0mm 5mm 0mm, clip ,width=0.95\columnwidth,draft=false]{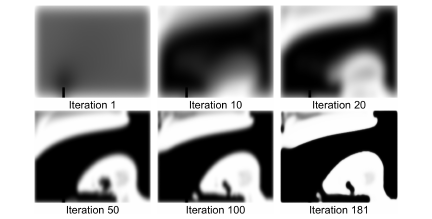}
\caption{}
\label{fig:Ant2Evolution}
\end{subfigure}
\\
    \centering
    \begin{subfigure}[t]{0.24\textwidth}
    \includegraphics[height=0.9\textwidth]{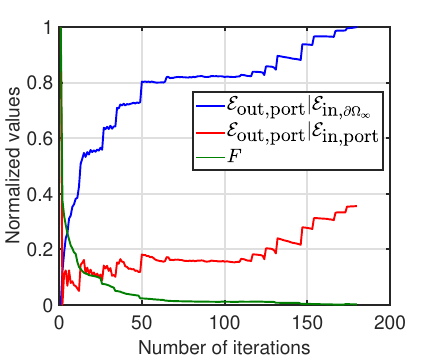}
    \caption{}
    \label{fig:Antobj}
    \end{subfigure}
            \centering
    \begin{subfigure}[t]{0.24\textwidth}
    \includegraphics[height=0.9\linewidth]{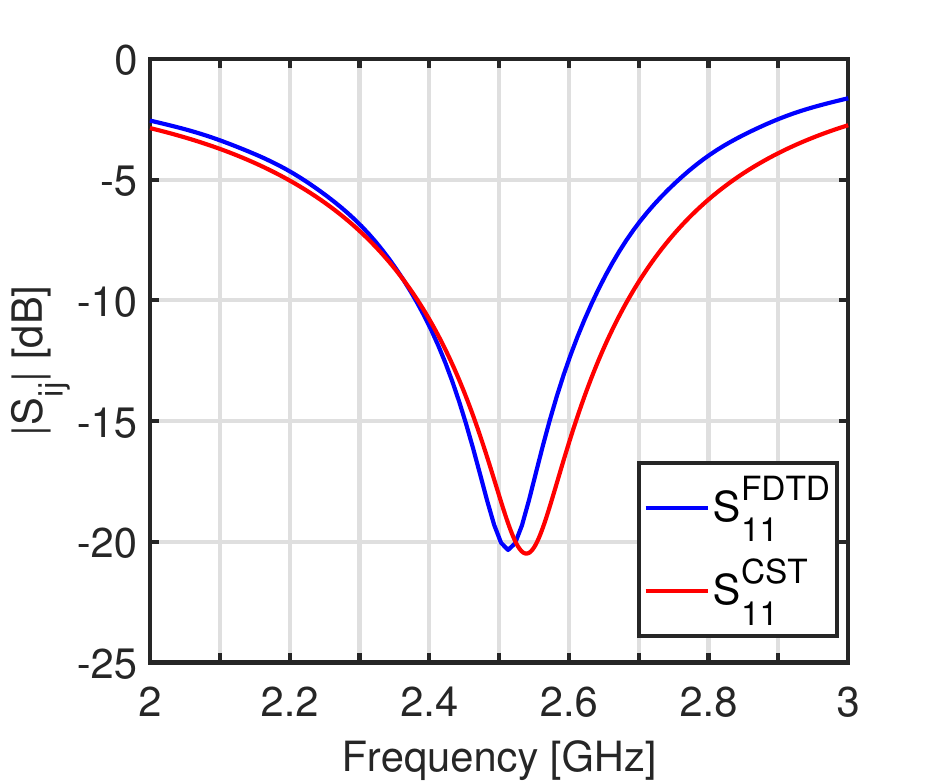}
    \caption{}
    \label{fig:AntSpara}
    \end{subfigure}
  
    \caption{Single antenna optimization. (a) Snapshots showing the development of the antenna structure during optimization. (b) Iteration history of the normalized objective function and its two sub-objectives. 
    (c) The reflection coefficient $S_{11}$ of the optimized antenna, computed with our FDTD code and compared with the CST Studio Suite.}
    \label{fig:antopt}
\end{figure}

The evolution of the antenna structure during the optimization process is shown in Fig.\,\ref{fig:Ant2Evolution}, where most of the intermediate values vanish gradually. 
After around 50 iterations, the overall structure is established, and the later iterations build small features to improve the performance, leading to convergence after 181 iterations. 
As shown in Fig.\,\ref{fig:Antobj}, the objective function $F$ evolves monotonically; however, the outgoing energies through the feeding  port, based on the transmitting ($\mathcal{E}_{\text{out,port}}\big|_{\mathcal{E}_{\text{in,port}}}$) or receiving ($\mathcal{E}_{\text{out,port}}\big|_{\mathcal{E}_{\text{in,$\partial\Omega\infty$}}}$) modes, experience fluctuations due to competition with each other. 
The objective functions suddenly change when the filter parameters are updated. 
The nonlinearity of the filter increases as $\alpha$ reduces to small values, resulting in less gray materials and an obvious increase of the terms involved in the objective function. 
The final design consists mainly of design variables close to 0, a good dielectric, or 1, a good conductor, besides some small regions near the boundaries with intermediate values. 
A thresholding value of 0.5 is used to map the final conductivities to be $0$ S/m and $5.8\times 10^7$ S/m.
The design from our FDTD models is exported as an STL file, which is then imported into CST microwave studio for cross-validation. 
The $|S_{11}|$ of the antenna, calculated using our FDTD code and the CST package, are presented in Fig.\,\ref{fig:AntSpara}, showing good agreement.
The slight differences can be attributed to the modeling discrepancies between the two simulation techniques. 
The optimized antenna achieves a good performance in the frequency range 2.4--2.6\,GHz with $|S_{11}|\approx-20$\,dB minimum at 2.5\,GHz.

\bibliographystyle{IEEEtran}
\bibliography{IEEEabrv,ref}

\begin{thebibliography}{10}
\providecommand{\url}[1]{#1}
\csname url@samestyle\endcsname
\providecommand{\newblock}{\relax}
\providecommand{\bibinfo}[2]{#2}
\providecommand{\BIBentrySTDinterwordspacing}{\spaceskip=0pt\relax}
\providecommand{\BIBentryALTinterwordstretchfactor}{4}
\providecommand{\BIBentryALTinterwordspacing}{\spaceskip=\fontdimen2\font plus
\BIBentryALTinterwordstretchfactor\fontdimen3\font minus \fontdimen4\font\relax}
\providecommand{\BIBforeignlanguage}[2]{{%
\expandafter\ifx\csname l@#1\endcsname\relax
\typeout{** WARNING: IEEEtran.bst: No hyphenation pattern has been}%
\typeout{** loaded for the language `#1'. Using the pattern for}%
\typeout{** the default language instead.}%
\else
\language=\csname l@#1\endcsname
\fi
#2}}
\providecommand{\BIBdecl}{\relax}
\BIBdecl

\bibitem{jha2018compact}
K.~R. Jha, B.~Bukhari, C.~Singh, G.~Mishra, and S.~K. Sharma, ``Compact planar multistandard {MIMO} antenna for {IoT} applications,'' \emph{{IEEE} Trans. Antennas Propag.}, vol.~66, no.~7, pp. 3327--3336, 2018.

\bibitem{sun2018compact}
L.~Sun, H.~Feng, Y.~Li, and Z.~Zhang, ``Compact {5G} {MIMO} mobile phone antennas with tightly arranged orthogonal-mode pairs,'' \emph{{IEEE} Trans. Antennas Propag.}, vol.~66, no.~11, pp. 6364--6369, 2018.

\bibitem{sun2020self}
L.~Sun, Y.~Li, Z.~Zhang, and H.~Wang, ``Self-decoupled {MIMO} antenna pair with shared radiator for {5G} smartphones,'' \emph{{IEEE} Trans. Antennas Propag.}, vol.~68, no.~5, pp. 3423--3432, 2020.

\bibitem{hei2021wideband}
Y.~Q. Hei, J.~G. He, and W.~T. Li, ``Wideband decoupled 8-element {MIMO} antenna for {5G} mobile terminal applications,'' \emph{IEEE Antennas Wirel. Propag. Lett.}, vol.~20, no.~8, pp. 1448--1452, 2021.

\bibitem{pan2021design}
Y.~M. Pan, Y.~Hu, and S.~Y. Zheng, ``Design of low mutual coupling dielectric resonator antennas without using extra decoupling element,'' \emph{{IEEE} Trans. Antennas Propag.}, vol.~69, no.~11, pp. 7377--7385, 2021.

\bibitem{liu2022mutual}
X.~Liu, S.~Gao, B.~Sanz-Izquierdo, H.~Zhang, L.~Wen, W.~Hu, Q.~Luo, J.~T.~S. Sumantyo, and X.-X. Yang, ``A mutual-coupling-suppressed dual-band dual-polarized base station antenna using multiple folded-dipole antenna,'' \emph{{IEEE} Trans. Antennas Propag.}, vol.~70, no.~12, pp. 11\,582--11\,594, 2022.

\bibitem{lai2021mutual}
Q.~X. Lai, Y.~M. Pan, S.~Y. Zheng, and W.~J. Yang, ``Mutual coupling reduction in {MIMO} microstrip patch array using {$\text{TM}_{10}$} and {$\text{TM}_ {02}$} modes,'' \emph{{IEEE} Trans. Antennas Propag}, vol.~69, no.~11, pp. 7562--7571, 2021.

\bibitem{li2021decoupling}
M.~Li, Y.~Zhang, D.~Wu, K.~L. Yeung, L.~Jiang, and R.~Murch, ``Decoupling and matching network for dual-band {MIMO} antennas,'' \emph{{IEEE} Trans. Antennas Propag.}, vol.~70, no.~3, pp. 1764--1775, 2021.

\bibitem{zhang2021simple}
Y.-M. Zhang, Q.-C. Ye, G.~F. Pedersen, and S.~Zhang, ``A simple decoupling network with filtering response for patch antenna arrays,'' \emph{{IEEE} Trans. Antennas Propag.}, vol.~69, no.~11, pp. 7427--7439, 2021.

\bibitem{pei2021low}
T.~Pei, L.~Zhu, J.~Wang, and W.~Wu, ``A low-profile decoupling structure for mutual coupling suppression in {MIMO} patch antenna,'' \emph{{IEEE} Trans. Antennas Propag.}, vol.~69, no.~10, pp. 6145--6153, 2021.

\bibitem{zhao2024three}
G.~Zhao, L.~Zhao, M.~Xi, Y.~Guo, G.-L. Huang, Y.~Li, X.~Wang, and W.~Lin, ``A three-port coupled resonator decoupling network for mutual coupling reduction of three-element antenna arrays,'' \emph{IEEE Trans. Microw. Theory Techn.}, 2024.

\bibitem{xu2020decoupling}
K.-D. Xu, H.~Luyen, and N.~Behdad, ``A decoupling and matching network design for single-and dual-band two-element antenna arrays,'' \emph{IEEE Trans. Microw. Theory Techn.}, vol.~68, no.~9, pp. 3986--3999, 2020.

\bibitem{khandelwal2017defected}
M.~K. Khandelwal, B.~K. Kanaujia, and S.~Kumar, ``Defected ground structure: fundamentals, analysis, and applications in modern wireless trends,'' \emph{Int. J. Antennas Propag.}, vol. 2017, no.~1, p. 2018527, 2017.

\bibitem{qian2021decoupling}
B.~Qian, X.~Chen, and A.~A. Kishk, ``Decoupling of microstrip antennas with defected ground structure using the common/differential mode theory,'' \emph{IEEE Antennas Wirel. Propag. Lett.}, vol.~20, no.~5, pp. 828--832, 2021.

\bibitem{wang2013neutralization}
Y.~Wang and Z.~Du, ``A wideband printed dual-antenna with three neutralization lines for mobile terminals,'' \emph{{IEEE} Trans. Antennas Propag.}, vol.~62, no.~3, pp. 1495--1500, 2013.

\bibitem{zhang2015line}
S.~Zhang and G.~F. Pedersen, ``Mutual coupling reduction for {UWB} {MIMO} antennas with a wideband neutralization line,'' \emph{{IEEE} Antennas Wirel. Propag. Lett.}, vol.~15, pp. 166--169, 2015.

\bibitem{Metasurface}
F.~Liu, J.~Guo, L.~Zhao, G.-L. Huang, Y.~Li, and Y.~Yin, ``Dual-band metasurface-based decoupling method for two closely packed dual-band antennas,'' \emph{{IEEE} Trans. Antennas Propag.}, vol.~68, no.~1, pp. 552--557, 2020.

\bibitem{Metasurface1}
H.~Luan, C.~Chen, W.~Chen, L.~Zhou, H.~Zhang, and Z.~Zhang, ``Mutual coupling reduction of closely {E/H}-plane coupled antennas through metasurfaces,'' \emph{{IEEE} Antennas Wirel. Propag. Lett.}, vol.~18, no.~10, pp. 1996--2000, 2019.

\bibitem{Liu24Full}
G.-Y. Liu, N.~Yang, K.~W. Leung, and K.~M. Luk, ``A full design perspective of port decoupling for {MIMO} antenna: Preservation of radiation pattern,'' \emph{{IEEE} Trans. Antennas Propag.}, pp. 1--1, 2024.

\bibitem{Koziel13Multi}
S.~Koziel and S.~Ogurtsov, ``Multi-objective design of antennas using variable-fidelity simulations and surrogate models,'' \emph{{IEEE} Trans. Antennas Propag}, vol.~61, no.~12, pp. 5931--5939, 2013.

\bibitem{Hassan19Compact}
E.~Hassan, D.~Martynenko, E.~Wadbro, G.~Fischer, and M.~Berggren, ``Compact differential-fed planar filtering antennas,'' \emph{Electronics}, vol.~8, no.~11, 2019.

\bibitem{Slawomir22Tolerance}
S.~Koziel and A.~Pietrenko-Dabrowska, ``Tolerance-aware multi-objective optimization of antennas by means of feature-based regression surrogates,'' \emph{{IEEE} Trans. Antennas Propag.}, vol.~70, no.~7, pp. 5636--5646, 2022.

\bibitem{cheng2023novel}
Y.-F. Cheng, D.~Li, S.~Chen, and G.~Wang, ``A novel wideband decoupling method based on even-odd-mode analysis and genetic algorithm optimization,'' \emph{{IEEE} Antennas Wirel. Propag. Lett.}, vol.~22, no.~10, pp. 2507--2511, 2023.

\bibitem{Hassan14patch}
E.~Hassan, E.~Wadbro, and M.~Berggren, ``Patch and ground plane design of microstrip antennas by material distribution topology optimization,'' \emph{Prog. Electromagn. Res. B}, vol.~59, p. 89{\textendash}102, 2014.

\bibitem{ko2024}
S.~Koziel and A.~Pietrenko-Dabrowska, ``Efficient simulation-based global antenna optimization using characteristic point method and nature-inspired metaheuristics,'' \emph{{IEEE} Trans. Antennas Propag}, vol.~72, no.~4, pp. 3706--3717, 2024.

\bibitem{Mousavi24Topology}
A.~Mousavi, M.~Berggren, L.~Hägg, and E.~Wadbro, ``Topology optimization of a waveguide acoustic black hole for enhanced wave focusing,'' \emph{J. Acoust. Soc. Am.}, vol. 155, no.~1, pp. 742--756, 01 2024.

\bibitem{Gedeon2023}
J.~Gedeon, E.~Hassan, and A.~Calà~Lesina, ``Time-domain topology optimization of arbitrary dispersive materials for broadband {3D} nanophotonics inverse design,'' \emph{ACS Photonics}, vol.~10, no.~11, pp. 3875--3887, 2023.

\bibitem{Hassan:22}
E.~Hassan and A.~{Cal\`{a} Lesina}, ``Topology optimization of dispersive plasmonic nanostructures in the time-domain,'' \emph{Opt. Express}, vol.~30, no.~11, pp. 19\,557--19\,572, 2022.

\bibitem{sigmund2013topology}
O.~Sigmund and K.~Maute, ``Topology optimization approaches: A comparative review,'' \emph{Struct. Multidiscip. Optim.}, vol.~48, no.~6, pp. 1031--1055, 2013.

\bibitem{Gedeon25Time}
J.~Gedeon, I.~Allayarov, A.~C. Lesina, and E.~Hassan, ``Time-domain topology optimization of power dissipation in dispersive dielectric and plasmonic nanostructures,'' \emph{{IEEE} Trans. Antennas Propag.}, vol.~73, no.~5, pp. 3079--3094, 2025.

\bibitem{liu2016mom}
S.~Liu, Q.~Wang, and R.~Gao, ``{MoM}-based topology optimization method for planar metallic antenna design,'' \emph{Acta Mechanica Sinica}, vol.~32, pp. 1058--1064, 2016.

\bibitem{hassan2018topology}
E.~Hassan, E.~Wadbro, L.~H{\"a}gg, and M.~Berggren, ``Topology optimization of compact wideband coaxial-to-waveguide transitions with minimum-size control,'' \emph{Struct. Multidiscip. Optim.}, vol.~57, pp. 1765--1777, 2018.

\bibitem{zhu2019design}
S.-H. Zhu, X.-S. Yang, J.~Wang, and B.-Z. Wang, ``Design of {MIMO} antenna isolation structure based on a hybrid topology optimization method,'' \emph{{IEEE} Trans. Antennas Propag.}, vol.~67, no.~10, pp. 6298--6307, 2019.

\bibitem{emad2020waveguide}
E.~Hassan, B.~Scheiner, F.~Michler, M.~Berggren, E.~Wadbro, F.~Röhrl, S.~Zorn, R.~Weigel, and F.~Lurz, ``Multilayer topology optimization of wideband {SIW}-to-waveguide transitions,'' \emph{{IEEE} Trans. Microw. Theory Tech.}, vol.~68, no.~4, pp. 1326--1339, 2020.

\bibitem{topothin}
T.~Jibiki, T.~Kawasaki, M.~Tanomura, and H.~Igarashi, ``Topology optimization of microwave devices with thin structure,'' \emph{{IEEE} Trans. Magn.}, pp. 1--1, 2024.

\bibitem{Tucek23Density}
J.~Tucek, M.~Capek, L.~Jelinek, and O.~Sigmund, ``Density-based topology optimization in method of moments: Q-factor minimization,'' \emph{{IEEE} Trans. Antennas Propag.}, vol.~71, no.~12, pp. 9738--9751, 2023.

\bibitem{smith2019gathin}
J.~S. Smith and M.~E. Baginski, ``Thin-wire antenna design using a novel branching scheme and genetic algorithm optimization,'' \emph{{IEEE} Trans. Antennas Propag.}, vol.~67, no.~5, pp. 2934--2941, 2019.

\bibitem{wang2017antenna}
J.~Wang, X.-S. Yang, X.~Ding, and B.-Z. Wang, ``Antenna radiation characteristics optimization by a hybrid topological method,'' \emph{{IEEE} Trans. Antennas Propag.}, vol.~65, no.~6, pp. 2843--2854, 2017.

\bibitem{hassan2014topology}
E.~Hassan, E.~Wadbro, and M.~Berggren, ``Topology optimization of metallic antennas,'' \emph{{IEEE} Trans. Antennas Propag.}, vol.~62, no.~5, pp. 2488--2500, 2014.

\bibitem{Pozar21microwave}
D.~Pozar, \emph{Microwave Engineering}, 4th~ed.\hskip 1em plus 0.5em minus 0.4em\relax John Wiley \& Sons, 2012.

\bibitem{cpml}
J.~A. Roden and S.~D. Gedney, ``Convolution {PML} ({CPML}): An efficient {FDTD} implementation of the {CFS--PML} for arbitrary media,'' \emph{Microw. Opt. Technol. Lett.}, vol.~27, no.~5, pp. 334--339, 2000.

\bibitem{taflove2005computational}
A.~Taflove, S.~C. Hagness, and M.~Piket-May, ``Computational electromagnetics: the finite-difference time-domain method,'' \emph{The Electrical Engineering Handbook}, vol.~3, no. 629-670, p.~15, 2005.

\bibitem{bokhari2023topology}
A.~H. Bokhari, E.~Hassan, and E.~Wadbro, ``Topology optimization of microwave frequency dividing multiplexers,'' \emph{Struct. Multidiscip. Optim.}, vol.~66, no.~5, p. 106, 2023.

\bibitem{clausen17OnFilter}
A.~Clausen and E.~Andreassen, ``\BIBforeignlanguage{en}{On filter boundary conditions in topology optimization},'' \emph{\BIBforeignlanguage{en}{Struct. Multidiscip. Optim.}}, vol.~56, no.~5, pp. 1147--1155, Nov. 2017.

\bibitem{HaWa16}
L.~H{\"a}gg and E.~Wadbro, ``Nonlinear filters in topology optimization: existence of solutions and efficient implementation for minimum compliance problems,'' \emph{Struct. Multidiscip. Optim.}, vol.~55, no.~3, p. 1017{\textendash}1028, 2017.

\bibitem{nikolova2006sensitivity}
N.~K. Nikolova, Y.~Li, Y.~Li, and M.~H. Bakr, ``Sensitivity analysis of scattering parameters with electromagnetic time-domain simulators,'' \emph{{IEEE} Trans. Microw. Theory Tech.}, vol.~54, no.~4, pp. 1598--1610, 2006.

\bibitem{hassan2015time}
E.~Hassan, E.~Wadbro, and M.~Berggren, ``Time-domain sensitivity analysis for conductivity distribution in {M}axwell's equations,'' Dept. of Computing Science, Ume{\aa} University, Tech. Rep. UMINF 15.06, 2015.

\bibitem{SvanbergGlobally}
K.~Svanberg, ``A class of globally convergent optimization methods based on conservative convex separable approximations,'' \emph{SIAM J. Optim.}, vol.~12, no.~2, p. 555{\textendash}573, 2002.

\bibitem{GPUFDTD}
P.~Lu and P.~Kosmas, ``Three-dimensional microwave head imaging with {GPU}-based {FDTD} and the {DBIM} method,'' \emph{Sensors}, vol.~22, no.~7, p. 2691, 2022.

\bibitem{harris2007optimizing}
M.~Harris, ``Optimizing parallel reduction in {CUDA},'' \emph{{Nvidia} developer technology}, vol.~2, no.~4, p.~70, 2007.

\bibitem{cst}
\BIBentryALTinterwordspacing
{Dassault Systèmes}, ``{CST Studio Suite},'' 2025, accessed on 06.02.2025. [Online]. Available: \url{https://www.3ds.com/products/simulia/cst-studio-suite}
\BIBentrySTDinterwordspacing

\bibitem{ding2020novel}
D.~Ding, D.~Li, J.~Xia, and Z.~Li, ``Novel optimization strategies for isolation structure design in {MIMO} systems,'' \emph{IEICE Electron. Express}, vol.~17, no.~8, pp. 1--4, 2020.

\bibitem{whaleswarm2025}
E.~Atashpanjeh and P.~Rezaei, ``Innovative design for mutual coupling reduction in dual-element array antennas for {ISM} applications using whale optimization algorithm,'' \emph{Prog. Electromagn. Res. C}, vol. 152, pp. 121--129, 2025.

\end{thebibliography}

\end{document}